\documentclass[11pt,letterpaper]{article}

\usepackage{fullpage}

\def\stocmode{0}

\def\arxivmode{0}
\def\fastmode{0}

\def\showauthornotes{2}

\def\showkeys{0}
\def\showdraftbox{1}
\def\showcolorlinks{1}
\def\usemicrotype{1}
\def\showfixme{1}

\newcommand{\bdelta}{\hat{\delta}}

\ifnum\arxivmode=0  \else  \fi

\ifnum\fastmode=0
\usepackage{etex}
\fi

\ifnum\fastmode=0
\usepackage[l2tabu, orthodox]{nag}
\fi

\usepackage{xspace,enumerate}

\usepackage[dvipsnames]{xcolor}

\ifnum\fastmode=0
\usepackage[T1]{fontenc}
\usepackage[full]{textcomp}
\fi

\usepackage[american]{babel}

\usepackage{mathtools}

\usepackage{amsthm}

\newtheorem{theorem}{Theorem}[section]
\newtheorem*{theorem*}{Theorem}

\newtheorem*{proposition*}{Proposition}
\newtheorem{lemma}[theorem]{Lemma}
\newtheorem*{lemma*}{Lemma}
\newtheorem{corollary}[theorem]{Corollary}
\newtheorem*{conjecture*}{Conjecture}
\newtheorem{fact}[theorem]{Fact}
\newtheorem*{fact*}{Fact}

\newtheorem*{exercise*}{Exercise}

\newtheorem*{hypothesis*}{Hypothesis}

\theoremstyle{definition}
\newtheorem{definition}[theorem]{Definition}

\newtheorem{example}[theorem]{Example}

\newtheorem{exercise-easy}[theorem]{Exercise}
\newtheorem{exercise-med}[theorem]{Exercise}
\newtheorem{exercise-hard}[theorem]{Exercise$^\star$}

\newtheorem*{claim*}{Claim}

\newtheorem{remark}[theorem]{Remark}
\newtheorem*{remark*}{Remark}

\newtheorem*{observation*}{Observation}

\ifnum\arxivmode=0
\usepackage{newpxtext} %
\usepackage{textcomp} %
\usepackage[varg,bigdelims]{newpxmath}
\usepackage[scr=boondoxo,scrscaled=1.10]{mathalfa}

\usepackage{bm} %
\let\mathbb\varmathbb
\fi

\ifnum\arxivmode=1
\usepackage[varg]{pxfonts} %
\usepackage{textcomp} %
\usepackage[scr=rsfso]{mathalfa}%
\usepackage{bm} %
\fi

\ifnum\showkeys=1
\usepackage[color]{showkeys}
\fi

\makeatletter
\@ifpackageloaded{hyperref}
{}
{\usepackage[pagebackref]{hyperref}}

\definecolor{bleudefrance}{rgb}{0.01, 0.1, 1.0}
\definecolor{azure}{rgb}{0.0, 0.5, 1.0}

\ifnum\showcolorlinks=1
\hypersetup{
colorlinks=true,
urlcolor=blue,
linkcolor=blue,
citecolor=OliveGreen}
\fi

\ifnum\showcolorlinks=0
\hypersetup{
pagebackref=true,
colorlinks=false,
pdfborder={0 0 0}}
\fi

\usepackage{prettyref}

\newcommand{\savehyperref}[2]{\texorpdfstring{\hyperref[#1]{#2}}{#2}}

\newrefformat{eq}{\savehyperref{#1}{\textup{(\ref*{#1})}}}
\newrefformat{lem}{\savehyperref{#1}{Lemma~\ref*{#1}}}
\newrefformat{def}{\savehyperref{#1}{Definition~\ref*{#1}}}
\newrefformat{obs}{\savehyperref{#1}{Observation~\ref*{#1}}}
\newrefformat{ass}{\savehyperref{#1}{Assumption~\ref*{#1}}}
\newrefformat{thm}{\savehyperref{#1}{Theorem~\ref*{#1}}}
\newrefformat{cor}{\savehyperref{#1}{Corollary~\ref*{#1}}}
\newrefformat{cha}{\savehyperref{#1}{Chapter~\ref*{#1}}}
\newrefformat{sec}{\savehyperref{#1}{Section~\ref*{#1}}}
\newrefformat{app}{\savehyperref{#1}{Appendix~\ref*{#1}}}
\newrefformat{tab}{\savehyperref{#1}{Table~\ref*{#1}}}
\newrefformat{fig}{\savehyperref{#1}{Figure~\ref*{#1}}}
\newrefformat{hyp}{\savehyperref{#1}{Hypothesis~\ref*{#1}}}
\newrefformat{alg}{\savehyperref{#1}{Algorithm~\ref*{#1}}}
\newrefformat{rem}{\savehyperref{#1}{Remark~\ref*{#1}}}
\newrefformat{item}{\savehyperref{#1}{Item~\ref*{#1}}}
\newrefformat{step}{\savehyperref{#1}{step~\ref*{#1}}}
\newrefformat{conj}{\savehyperref{#1}{Conjecture~\ref*{#1}}}
\newrefformat{fact}{\savehyperref{#1}{Fact~\ref*{#1}}}
\newrefformat{prop}{\savehyperref{#1}{Proposition~\ref*{#1}}}
\newrefformat{prob}{\savehyperref{#1}{Problem~\ref*{#1}}}
\newrefformat{claim}{\savehyperref{#1}{Claim~\ref*{#1}}}
\newrefformat{relax}{\savehyperref{#1}{Relaxation~\ref*{#1}}}
\newrefformat{red}{\savehyperref{#1}{Reduction~\ref*{#1}}}
\newrefformat{part}{\savehyperref{#1}{Part~\ref*{#1}}}
\newrefformat{ex}{\savehyperref{#1}{Exercise~\ref*{#1}}}
\newrefformat{property}{\savehyperref{#1}{Property~\ref*{#1}}}

\newcommand{\Sref}[1]{\hyperref[#1]{\S\ref*{#1}}}

\usepackage{nicefrac}

\ifnum\fastmode=0
\ifnum\usemicrotype=1
\usepackage{microtype}
\fi
\fi

\ifnum\showauthornotes=2
\newcommand{\mynotes}[1]{{\sffamily\small\color{teal}{#1}}\medskip}
\newcommand{\Authornote}[2]{{\sffamily\small\color{Maroon}{[#1: #2]}}\medskip}
\newcommand{\Authornotecolored}[3]{{\sffamily\small\color{#1}{[#2: #3]}}}
\newcommand{\Authorcomment}[2]{{\sffamily\small\color{gray}{[#1: #2]}}}
\newcommand{\Authorstartcomment}[1]{\sffamily\small\color{gray}[#1: }

\newcommand{\Authorfnote}[2]{\footnote{\color{red}{#1: #2}}}
\newcommand{\Authorfixme}[1]{\Authornote{#1}{\textbf{??}}}
\newcommand{\Authormarginmark}[1]{\marginpar{\textcolor{red}{\fbox{\Large #1:!}}}}
\newcommand{\myexplain}[1]{{\sffamily\small\color{red}{\noindent [Explanation:\medskip\newline \begin{quote}#1\hfill]\end{quote}}}\medskip}
\newcommand{\explain}[1]{{\sffamily\small\color{red}{#1}}\medskip}

\else

\newcommand{\mynotes}[1]{}
\newcommand{\Authornote}[2]{}
\newcommand{\Authornotecolored}[3]{}
\newcommand{\Authorcomment}[2]{}
\newcommand{\Authorstartcomment}[1]{}

\newcommand{\Authorfnote}[2]{}
\newcommand{\Authorfixme}[1]{}
\newcommand{\Authormarginmark}[1]{}
\newcommand{\myexplain}[1]{}
\newcommand{\explain}[1]{}

\ifnum\showauthornotes=1
\renewcommand{\myexplain}[1]{{\sffamily\small\color{red}{\noindent \begin{quote}{\bf Explanation:} \medskip\newline #1\end{quote}}}\medskip}
\fi

\fi

\ifnum\showfixme=0

\fi

\usepackage{boxedminipage}

\newcommand{\norm}[1]{\lVert#1\rVert}

\newcommand{\Esymb}{\mathbb{E}}
\newcommand{\Psymb}{\mathbb{P}}

\DeclareMathOperator*{\E}{\Esymb}

\DeclareMathOperator*{\ProbOp}{\Psymb}

\renewcommand{\Pr}{\ProbOp}

\newcommand{\textparen}[1]{\text{(#1)}}

\ifx\because\undefined
\newcommand{\because}[1]{\textparen{because #1}}
\else
\renewcommand{\because}[1]{\textparen{because #1}}
\fi

\newcommand{\seteq}{\mathrel{\mathop:}=}

\newcommand\bdot\bullet

\ifx\mathds\undefined %

\else

\fi

\DeclareMathOperator{\vol}{vol}
\DeclareMathOperator{\poly}{poly}

\DeclareMathOperator{\supp}{supp}

\DeclareMathOperator{\sign}{sign}

\DeclareMathOperator{\rank}{rank}

\newcommand{\Z}{\mathbb Z}

\newcommand{\R}{\mathbb R}

\newcommand{\cA}{\mathcal A}

\newcommand{\cC}{\mathcal C}

\newcommand{\cL}{\mathcal L}

\newcommand{\cN}{\mathcal N}

\newcommand{\cS}{\mathcal S}

\renewcommand{\leq}{\leqslant}
\renewcommand{\le}{\leqslant}
\renewcommand{\geq}{\geqslant}
\renewcommand{\ge}{\geqslant}

\ifnum\showdraftbox=1

\else

\fi

\let\epsilon=\varepsilon

\numberwithin{equation}{section}

\newcommand\MYcurrentlabel{xxx}

\newcommand{\MYstore}[2]{%
  \global\expandafter \def \csname MYMEMORY #1 \endcsname{#2}%
}

\newcommand{\MYload}[1]{%
  \csname MYMEMORY #1 \endcsname%
}

\newcommand{\MYnewlabel}[1]{%
  \renewcommand\MYcurrentlabel{#1}%
  \MYoldlabel{#1}%
}

\newcommand{\MYdummylabel}[1]{}

\newcommand{\torestate}[1]{%
  \let\MYoldlabel\label%
  \let\label\MYnewlabel%
  #1%
  \MYstore{\MYcurrentlabel}{#1}%
  \let\label\MYoldlabel%
}

\newcommand{\restatetheorem}[1]{%
  \let\MYoldlabel\label
  \let\label\MYdummylabel
  \begin{theorem*}[Restatement of \prettyref{#1}]
    \MYload{#1}
  \end{theorem*}
  \let\label\MYoldlabel
}

\newcommand{\restatelemma}[1]{%
  \let\MYoldlabel\label
  \let\label\MYdummylabel
  \begin{lemma*}[Restatement of \prettyref{#1}]
    \MYload{#1}
  \end{lemma*}
  \let\label\MYoldlabel
}

\newcommand{\restateprop}[1]{%
  \let\MYoldlabel\label
  \let\label\MYdummylabel
  \begin{proposition*}[Restatement of \prettyref{#1}]
    \MYload{#1}
  \end{proposition*}
  \let\label\MYoldlabel
}

\newcommand{\restatefact}[1]{%
  \let\MYoldlabel\label
  \let\label\MYdummylabel
  \begin{fact*}[Restatement of \prettyref{#1}]
    \MYload{#1}
  \end{fact*}
  \let\label\MYoldlabel
}

\newcommand{\restate}[1]{%
  \let\MYoldlabel\label
  \let\label\MYdummylabel
  \MYload{#1}
  \let\label\MYoldlabel
}

\newcommand{\addreferencesection}{
  \phantomsection
\ifnum\stocmode=0
  \addcontentsline{toc}{section}{References}
\else
  \addcontentsline{toc}{section}{References \hspace*{1in} --------- End of extended abstract ---------}
\fi

}

\newcommand{\e}{\epsilon}
\newcommand{\eps}{\epsilon}

\renewcommand{\paragraph}[1]{\medskip\noindent{\bf #1.}}

\allowdisplaybreaks

\usepackage{paralist}

\usepackage{comment}
\usepackage[algo2e,linesnumbered,lined,ruled,boxed]{algorithm2e}
\usepackage{algpseudocode}

\definecolor{orange}{rgb}{0.8, 0.33, 0.0}

\let\pref=\prettyref

\newcommand{\cov}{\mathcal{K}}
\newcommand{\diam}{\mathrm{diam}}

\DeclareMathOperator{\tr}{tr}

\newcommand{\vertiii}[1]{{\left\vert\kern-0.25ex\left\vert\kern-0.25ex\left\vert #1 
          \right\vert\kern-0.25ex\right\vert\kern-0.25ex\right\vert}}

\newcommand\f{\varphi}

\newcommand{\tEval}{\mathcal{T}_{\mathrm{eval}}}

\usepackage{accents}

\usepackage{stmaryrd}

\usepackage{enumitem}

\renewcommand{\mathbb}{\vvmathbb}

\renewcommand{\supp}{\mathrm{supp}}

\begin{document}

\title{Sparsifying sums of norms}
\author{Arun Jambulapati\thanks{University of Washington, {\tt jmblpati@uw.edu}} \and 
James R. Lee\thanks{University of Washington, {\tt jrl@cs.washington.edu}} \and 
Yang P. Liu\thanks{Institute for Advanced Study, {\tt yangpliu@ias.edu}} \and
Aaron Sidford\thanks{Stanford University, {\tt sidford@stanford.edu}}}
\date{}

\maketitle

\begin{abstract}
   For any norms 	$N_1,\ldots,N_m$ on $\mathbb{R}^n$ and $N(x) :=
   N_1(x)+\cdots+N_m(x)$, we show there is a sparsified norm $\tilde{N}(x) = w_1
   N_1(x) + \cdots + w_m N_m(x)$ such that $|N(x) - \tilde{N}(x)| \leq \epsilon
   N(x)$ for all $x \in \mathbb{R}^n$, where $w_1,\ldots,w_m$ are non-negative
   weights, of which only $O(\epsilon^{-2} n \log(n/\epsilon) (\log n)^{2.5} )$ are
   non-zero.

   Additionally, if $N$ is $\mathrm{poly}(n)$-equivalent to the Euclidean norm on $\R^n$, then
   such weights can be found with high probability in time $O(m (\log n)^{O(1)} + \mathrm{poly}(n)) T$, where $T$ is the time required
  to evaluate a norm $N_i$. This immediately yields analogous statements for sparsifying sums of symmetric submodular functions.
  More generally, we show how to sparsify sums of $p$th powers of norms when the sum is $p$-uniformly smooth.
\end{abstract} 

\begingroup
\hypersetup{linktocpage=false}
\setcounter{tocdepth}{2}
\tableofcontents
\endgroup

\newpage

\section{Introduction}
\label{sec:intro}

Consider a collection $N_1,\ldots,N_m : \R^n \to \R_+$ of semi-norms\footnote{A semi-norm $N$ is nonnegative 
and satisfies $N(\lambda x) = |\lambda| N(x)$ and $N(x+y) \leq N(x)+N(y)$ for all $\lambda\in \R$, $x,y \in \R^n$, though possibly $N(x)=0$ for $x \neq 0$.}
on $\R^n$ and the semi-norm defined by
\[
   N(x) \seteq N_1(x)+\cdots+N_m(x)\,.
\]

It is natural to ask whether $N$ can be {\em sparsified} in the following sense.
Given nonnegative weights $w_1,\ldots,w_m$, define the approximator
$\tilde{N}(x) \seteq w_1 N_1(x) + \cdots + w_m N_m(x)$.
Say that $\tilde{N}$ is {\em $s$-sparse} if at most $s$ of the weights $\{w_i\}$ are non-zero,
and that $\tilde{N}$ is an {\em $\e$-approximation of $N$} if it holds that
\begin{equation}\label{eq:eps-approximator}
   |N(x)-\tilde{N}(x)| \leq \e N(x)\,,\quad \forall x\in \R^n\,.
\end{equation}

A prototypical example occurs for cut sparsifiers of weighted graphs.
In this case, one has an undirected graph $G=(V,E,c)$ with nonnegative weights $\{c_e : e \in E\}$, with
$n = |V|$ and
$N(x) \seteq \sum_{uv \in E} c_{uv} |x_u-x_v|$.
A {\em weighted cut sparsifier} is given by nonnegative edge weights $\{w_e : e\in E \}$.
Defining $\tilde{N}(x) \seteq \sum_{uv \in E} w_{uv} c_{uv} |x_u-x_v|$, the typical approximation
criterion is that
\begin{equation}\label{eq:cut-sparsifier}
   |N(x)-\tilde{N}(x)| \leq \e N(x)\,,\quad \forall x\in \{0,1\}^V\,,
\end{equation}
where $x \in \{0,1\}^V$ naturally indexes cuts in $G$.
A straightforward $\ell_1$ variant of the discrete Cheeger inequality 
shows that \eqref{eq:cut-sparsifier} is equivalent to \eqref{eq:eps-approximator}
in the setting of weighted graphs.

Bencz\'{u}r and Karger \cite{BK96} showed that for every graph $G$ and every $\e > 0$, one
can construct an $s$-sparse $\e$-approximate cut sparsifier with $s \leq O(\e^{-2} n \log n)$.
Their result addresses the case when each $N_i$ is a $1$-dimensional semi-norm of the form $N_i(x) = c_{uv} |x_u-x_v|$.
We show that one can obtain similar sparsifiers in substantial generality.

Further, we show how to compute such sparsifiers efficiently when the semi-norm $N$ is appropriately well-conditioned.
Say that $N$ is {\em $(r,R)$-rounded} if it holds that $r \|x\|_2 \leq N(x) \leq R \|x\|_2$ for all
$x \in \ker(N)^{\perp}$, where $\ker(N) \seteq \{ x \in \R^n : N(x) = 0 \}$.

\begin{theorem}
   \label{thm:l1-sparsification}
   Consider a collection $N_1,\ldots,N_m$ of semi-norms on $\R^n$ and $N(x) \seteq N_1(x) +\cdots+N_m(x)$.
   For every $\e > 0$, there is an $O(\e^{-2} n \log(n/\e) (\log n)^{2.5})$-sparse
   $\e$-approximation of $N$.  Further, if the semi-norm $N$ is $(r, R)$-rounded,
   then weights realizing the approximation can be found in time $O(m (\log
   n)^{O(1)} + n^{O(1)}) (\log(mR/r))^{O(1)} \tEval$ with high probability if each $N_i$
   can be evaluated in time $\tEval$.
\end{theorem}

\paragraph{Application to symmetric submodular functions}
A function $f : 2^V \to \R_{+}$ is {\em submodular} if 
\[
   f(S \cup \{v\}) - f(S) \geq f(T \cup \{v\}) - f(T),\quad
   \forall S \subseteq T \subseteq V, v \in V \setminus T\,.
\]
A submodular function is {\em symmetric} if  $f(S) = f(V\setminus S)$ for all $S \subseteq V$.

Consider submodular functions $f_1,\ldots,f_m : \{0,1\}^V \to \R_{+}$ and denote $F(S) \seteq f_1(S) + \cdots + f_m(S)$.
Given nonnegative weights $w_1,\ldots,w_m$, define $\tilde{F}(S) \seteq w_1 f_1(S) + \cdots + w_m f_m(S)$.
We say that $\tilde{F}$ is an {\em $s$-sparse $\e$-approximation for $F$} if it holds that
at most $s$ of the weights $\{w_i\}$ are non-zero and 
\[
   |F(S) - \tilde{F}(S)| \leq \e F(S)\,,\quad \forall S \subseteq V\,.
\]

Motivated by the ubiquity of submodular functions in machine learning and data mining,
Rafiey and Yoshida \cite{RY22} established in this setting that, even if the $f_i$ are asymmetric, for every $\e > 0$, there 
is an $O(Bn^2/\e^2)$-sparse $\e$-approximation for $F$, where $n \seteq |V|$ and $B$ is the maximum number of vertices in the base polytope of any $f_i$. In the case $B \le O(1)$, their result is tight for (directed) cuts in directed graphs \cite{CKPPRSV17}.

However, for {\em symmetric} submodular functions, the situation is better. For such functions $f : 2^V \to \R_+$ with $f(\emptyset) = 0$, the Lov\'asz extension \cite{Lovasz83} 
of $f$ is a semi-norm on $\R^V$ (see \pref{sec:sfm}). Therefore, \pref{thm:l1-sparsification} immediately yields an analogous sparsification result in this setting.
In comparison to \cite{RY22}, in this symmetric setting, we have no dependence on $B$, and the quadratic dependence on $n$ improves to nearly linear.

\begin{corollary}[Symmetric submodular functions]
\label{cor:sfm}
   If $f_1,\ldots,f_m : 2^V \to \R_{+}$ are symmetric submodular functions with $f_1(\emptyset) = \cdots = f_m(\emptyset) = 0$, and $F(S) \seteq f_1(S) + \cdots + f_m(S)$, then
   for every $\e > 0$, there is an 
   $O(\e^{-2} n \log(n/\e) (\log n)^{2.5})$-sparse
   $\e$-approximation of $F$, where $n = |V|$.
   
   Additionally, if the functions $f_i$ are integer-valued with $\max_{i \in [m], S \subseteq V} f_i(S) \le R$, then the weights realizing the approximation can be found in time $O(mn(\log n)^{O(1)} + \poly(n))\tEval\log^{O(1)}(mR)$, with high probability, assuming each $f_i$ can be evaluated in time $\tEval$. 
\end{corollary}

The deduction of \pref{cor:sfm} from \pref{thm:l1-sparsification} appears in \pref{sec:sfm}.

\paragraph{Sums of higher powers}
In the setting of graphs,  {\em spectral sparsification} \cite{ST11}, a notion stronger than \eqref{eq:cut-sparsifier}, has been extensively studied. 
Given semi-norms $N_1,\ldots,N_m$ on $\R^n$, define a semi-norm via their $\ell_2$-sum as 
\[
   N(x)^2 \seteq N_1(x)^2 + \cdots + N_m(x)^2\,.
\]
If $w_1,\ldots,w_m$ are nonnegative weights and $\tilde{N}(x)^2 \seteq w_1 N_1(x)^2 + \cdots + w_m N_m(x)^2$,
we say that $\tilde{N}^2$ is an {\em $s$-sparse $\e$-approximation for $N^2$} if it holds that
at most $s$ of the weights $\{w_i\}$ are non-zero and
\begin{equation}\label{eq:spectral-sparsifier}
   \left|N(x)^2 - \tilde{N}(x)^2\right| \leq \e N(x)^2\,,\quad \forall x \in \R^n\,.
\end{equation}

When $G=(V,E,c)$ is a weighted graph and each $N_i(x)$ is of the form $\sqrt{c_{uv}} |x_u-x_v|$ for some $uv \in E$,
\eqref{eq:spectral-sparsifier} is called an \emph{$\e$-spectral sparsifier of $G$}.
In this setting, a sequence of works \cite{ST11,SS11,BSS12} culminates in the existence
of $O(n/\e^2)$-sparse $\e$-approximations for every $\e > 0$.
These results generalize \cite{Rudelson99b, BSS14b} to the setting of arbitrary $1$-dimensional semi-norms,
where 
\begin{equation}\label{eq:one-d}
   N_1(x) =|\langle a_1,x\rangle|, \ldots, N_m(x) = |\langle a_m,x\rangle|,\qquad a_1,\ldots,a_m \in \R^n\,.
\end{equation}

We establish the existence of near-linear-size sparsifiers for sums of powers of a substantially more general class of higher-dimensional norms.
Recall that a semi-norm $N$ on $\R^n$ is said to be {\em $p$-uniformly smooth with constant $S$} 
if it holds that
\begin{equation}\label{eq:p-smooth}
   \frac{N(x+y)^p + N(x-y)^p}{2} \leq N(x)^p +  N(S y)^p,\qquad x,y \in \R^n\,.
\end{equation}
Note that when $N_i(x) = |\langle a_i, x \rangle|$, then $N$ is $2$-uniformly smooth with constant $1$.
We say that two semi-norms $N_X$ and $N_Y$ are {\em $K$-equivalent} if there is a number $\lambda > 0$
such that $N_Y(z) \leq \lambda N_X(z) \leq K N_Y(z)$ for all $z \in \R^n$.
Every norm is $1$-uniformly smooth with constant $1$ by the triangle inequality, so
the next theorem generalizes \pref{thm:l1-sparsification}.

\begin{theorem}[Sums of $p$th powers of uniformly smooth norms]
   \label{thm:l2-sparsification}
   Consider $p \geq 1$ and semi-norms $N_1,\ldots,N_m$ on $\R^n$.
   Denote $N(x)^p \seteq N_1(x)^p + \cdots + N_m(x)^p$, and
   suppose that for some numbers $K,S > 1$ the semi-norm $N$ is $K$-equivalent to a semi-norm which is $\min(p,2)$-uniformly smooth with constant $S$.
   Then for every $\e \in (0,1)$, there is an $O(s)$-sparse $\e$-approximation to $N^p$ such that
   \[
      s \leq \begin{cases}
         \frac{K^p}{\e^{2}} n \left(S \psi_n \log(n/\e)\right)^{p} (\log n)^{2} & 1 \leq p \leq 2 \\
      \frac{K^p S^p p^2}{\e^2} \left(\frac{n+p}{2}\right)^{p/2} \left(\psi_n \log(n/\e)\right)^2 (\log n)^2 & p \geq 2\,.
      \end{cases}
   \]
   Above, we use $\psi_n \leq O(\sqrt{\log n})$ \cite{Klartag23} to denote the KLS constant on $\R^n$
(see \pref{thm:kls-concentration} below).
\end{theorem}
Note that for $N(x)$ to be $\min(p,2)$-uniformly smooth with constant $O(S)$, it suffices that each $N_i$
is $\min(p,2)$-uniformly smooth with constant $S$ \cite{Figiel76}.
To see the relevance of this theorem in the case $p = 2$, note that by John's theorem, every $d$-dimensional semi-norm is $\sqrt{d}$-equivalent
to a Euclidean norm (which is $2$-uniformly smooth with constant $1$). So if $A_1,\ldots,A_m \in \R^{d \times n}$ and
$\hat{N}_1,\ldots,\hat{N}_m$ are {\em arbitrary} norms on $\R^d$, then taking
$N_i(x) \seteq \hat{N}_i(A_i x)$, we obtain an $O(d \e^{-2} n (\log(n/\e))^2 (\log n)^3)$-sparse $\e$-approximation to $N^2$,
substantially generalizing the setting of \eqref{eq:one-d} (albeit with an extra $d (\log (n/\e))^{O(1)}$ factor in the sparsity).

\smallskip

   Unlike in the setting of graph sparsifiers where spectral sparsification is a strictly stronger notion 
   (due to the equivalence of \eqref{eq:cut-sparsifier} and \eqref{eq:eps-approximator}), the notions of
   approximation guaranteed by \pref{thm:l1-sparsification} and \pref{thm:l2-sparsification} for $p > 1$ are, in general,
   incomparable. For example, even if $\|\tilde{A}x\|_2 \approx \|Ax\|_2$ for all $x \in \R^n$, it is not necessarily true that
   $\|\tilde{A}x\|_1 \approx \|Ax\|_1$ for all $x \in \R^n$.

Let us now discuss some consequences of \pref{thm:l2-sparsification}.

\paragraph{Dimension reduction for $\ell_p$ sums}
Fix $1 \leq p \leq 2$ and a subspace $X \subseteq \ell_p^m$ with $\dim(X)=n$. 
It is known that for any $\e > 0$, there is a subspace $\tilde{X} \subseteq \ell_p^d$
with $d \leq O(\e^{-2} n \log(n) (\log \log n)^2)$ such that
the $\ell_p$ norms on $X$ and $\tilde{X}$ are $(1+\e)$-equivalent \cite{Talagrand95}.
For $p=1$, this can be improved to $d \leq O(\e^{-2} n \log n)$ \cite{Talagrand90}.

Consider the following more general setting.
Suppose $Z_1,\ldots,Z_m$ are each $p$-uniformly smooth Banach spaces with
their smoothness constants bounded by $S$.
Let us write $(Z_1 \oplus \cdots \oplus Z_m)_p$ for the Banach space $Z = Z_1 \oplus \cdots \oplus Z_m$
equipped with the norm
\[
   \|x\|_{Z} \seteq \left(\|x\|_{Z_1}^p + \cdots + \|x\|_{Z_m}^p\right)^{1/p}.
\]
\pref{thm:l2-sparsification} shows the following:
For any $n$-dimensional subspace $X \subseteq Z$ and $\e > 0$,
there are indicies $i_1,\ldots,i_d \in \{1,\ldots,m\}$  with $d \leq O((S/\e)^{-2} n (\log (n/\e))^p (\log n)^{2+p/2})$
and a subspace
$\tilde{X} \subseteq (Z_{i_1} \oplus \cdots \oplus Z_{i_d})_p$
that is $(1+\e)$-equivalent to $X$.
The aforementioned results for subspaces of $\ell_p^m$ correspond to the setting where each $Z_i$ is $1$-dimensional.
The case $p \geq 2$ of \pref{thm:l2-sparsification} similarly generalizes \cite{BLM89}.

\paragraph{Application to spectral hypergraph sparsifiers}
Consider a weighted hypergraph $H=(V,E,c)$, where $\{ c_e : e \in E\}$ are nonnegative weights.
To every hyperedge $e \in E$, one can associate the semi-norm
$N_e(x) \seteq \sqrt{c_e} \max_{u,v \in e} |x_u-x_v|$, and the hypergraph energy
\[
   N(x)^2 \seteq \sum_{e \in E} N_e(x)^2\,.
\]

Soma and Yoshida \cite{SY19} formalized the notion of spectral sparsification for hypergraphs;
it coincides with the notion of approximation expressed in \eqref{eq:spectral-sparsifier}.
In this setting, a sequence of works \cite{SY19,BST19,KKTY21a,KKTY21b,JLS23,Lee23} culminates in the existence
of $O(\e^{-2} n (\log n)^{2})$-sparse
$\e$-approximations to $N^2$ for every $\e > 0$.

One can obtain a similar result via an application of \pref{thm:l2-sparsification}, as follows.
We can express each hyperedge norm as $N_e(x) = \|A_e x\|_{\infty}$, where $A_e : \R^n \to \R^{{|e| \choose 2}}$
is defined by $(A_e x)_{uv} = x_u - x_v$ for all $\{u,v\} \in {e \choose 2}$.
The $\ell_{\infty}$ norm on $\R^d$ is $K$-equivalent to the $\ell_{\lceil \log d \rceil}$ norm with $K = O(1)$, and the $\ell_p$ norm
on $\R^n$ is $2$-uniformly smooth with constant $S \leq O(\sqrt{p})$ \cite{Hanner56}.
Applying \pref{thm:l2-sparsification} with $S \leq O(\sqrt{\log n})$ and $K \leq O(1)$
yields $O(\e^{-2} n (\log (n/\e))^2 (\log n)^{4})$-sparse $\e$-approximators in this special case, nearly matching the known results on spectral hypergraph sparsification.
Additionally, \pref{thm:l2-sparsification} can be applied to give nontrivial sparsification results in substantially more general settings, as the next example shows.

\begin{example}[Sparsification for matrix norms]
Consider a matrix generalization of this setting:
$X \in \R^{d \times d}$, and matrices $S_1, \dots, S_m$ with $S_i \in \R^{d_i \times d}$, and $T_1, \dots, T_m$ with $T_i \in \R^{d \times e_i}$. Define
$N_i(X) \seteq \|S_i X T_i\|_{op}$, where $\|\cdot\|_{op}$ denotes the operator norm.
Then the semi-norm given by $N(X) \seteq (\|S_i X T_i\|_{op}^2 + \cdots + \|S_m X T_m\|_{op}^2)^{1/2}$ can be 
sparsified down to $O((d/\e)^2 (\log(d/\e))^2 (\log d)^4)$ terms.
This follows because the Schatten $p$-norm of an operator 
is $2$-uniformly smooth with constant $O(\sqrt{p})$ \cite{BCL94}, and
for rank $d$ matrices, the Schatten $p$-norm is $O(1)$-equivalent to the operator norm when $p \asymp \log d$.
\end{example}

\paragraph{Further results and open questions for sums of squared norms}
The {\em rank of a hypergraph $H$} is defined as the quantity $r \seteq \max_{e \in E} |e|$.
The best-known result for spectral hypergraph sparsification is due to 
\cite{JLS23,Lee23}: For every $\e > 0$, there is an $O(\e^{-2} \log(r) \cdot n \log n)$-sparse
$\e$-approximation to $N^2$.
In \pref{sec:sumsmooth}, we obtain the following generalization.

\begin{theorem}[Sums of squares of $\ell_p$ norms]
   \label{thm:sos-lp}
   Consider a family of linear operators $\{ A_i : \R^n \to \R^{k_i} \}_{i \in [m]}$, and $2 \le p_1, \dots, p_m \le p$.
   Suppose that $N_1,\ldots,N_m$ are semi-norms on $\R^n$ and that
   $N_i(x)$ is $K$-equivalent to $\|A_i x\|_{p_i}$ for all $i \in [m]$.
   Then for every $\e > 0$, there is an $O((K^3/\e)^2 p n \log(n/\e))$-sparse $\e$-approximation to $N^2$ where $N(x)^2 \seteq N_1(x)^2 + \cdots + N_m(x)^2$.
\end{theorem}

In particular, if $k_1,\ldots,k_m \leq r$, then each $\|A_i x\|_{\infty}$ is $O(1)$-equivalent to $\|A_i x\|_{p}$
for $p \asymp \log r$, and thus \pref{thm:sos-lp} generalizes the aforementioned result for spectral hypergraph sparsifiers.

\begin{corollary}\label{cor:rank-r}
   Consider a family of linear operators $\{ A_i : \R^n \to \R^{k_i} \}_{i \in [m]}$
   with $k_1,\ldots,k_m \leq r$, and define the semi-norm $N$ on $\R^n$ by
   \[
      N(x)^2 \seteq \|A_1 x\|_{\infty}^2 + \cdots + \|A_m x\|_{\infty}^2\,.
   \]
   Then for every $\e > 0$, there is an $O(\e^{-2} \log(r) \cdot n \log(n/\e))$-sparse $\e$-approximation to $N^2$.
\end{corollary}

One should note that, for any fixed $p \geq 2$, \pref{thm:sos-lp} is tight for methods based on independent sampling, by the
coupon collector bound.
(Although it is known that in some settings \cite{BSS12} the $\log(n)$ factor can be removed by other methods.)

It is a fascinating open question whether the assumption of $p$-uniform smoothness can be dropped from \pref{thm:l2-sparsification}.
In \pref{sec:sumsquare}, we show that it is possible to obtain a non-trivial result for sums of $p$th powers of general norms for $p \in [1,2]$.

\begin{theorem}[General sums of $p$th powers]
   \label{thm:sos-general}
   If $N_1,\ldots,N_m$ are arbitrary semi-norms on $\R^n$, $1 \leq p \leq 2$, and $N(x)^p \seteq N_1(x)^p + \cdots + N_m(x)^p$, then for every $\e > 0$, there is an $s$-sparse $\e$-approximation to $N^p$ with
   \[
      s \lesssim \e^{-2} \left(n^{2-1/p} \log(n/\e) (\log n)^{1/2} + n \log (n/\e)^p (\log n)^{2+p/2} \right)\,.
   \]
\end{theorem}

Note that in the $p = 2$ case, one obtains $s \leq O(\e^{-2}n^{3/2}\log(n/\e)^2 (\log n)^3)$.
This implies that the bound of \pref{cor:rank-r} cannot be sharp for $\log(r) \gg \sqrt{n}$,
as every $n$-dimensional normed space is $O(1)$-equivalent to a subspace of $\ell_{\infty}^{C^n}$ for some $C > 1$
(a proof of this standard fact occurs in \pref{sec:general-infinity}).

\subsection{Importance sampling for general norms}

Let us now fix semi-norms $N_1,\ldots,N_m$ on $\R^n$ and define $N(x) \seteq N_1(x) + \cdots + N_m(x)$ for all $x \in \R^n$, as in the setting of \pref{thm:l1-sparsification}.
Our method for constructing sparsifiers is simply independent sampling: Consider a probability distribution $\rho=(\rho_1,\ldots,\rho_m) \in (0,1]^m$ on $\{1,\ldots,m\}$,
and then sample $M$ indices $i_1,\ldots,i_M$ independently from $\rho$ and take
\[
   \tilde{N}(x) \seteq \frac{1}{M} \left(\frac{N_{i_1}(x)}{\rho_{i_1}} + \cdots + \frac{N_{i_m}(x)}{\rho_{i_m}}\right)\,.
\]
We have $\E[N_{i_1}(x)/\rho_{i_1}] = N(x)$, and therefore $\E[\tilde{N}(x)]=N(x)$ for any fixed $x$.

In order for these unbiased estimators to be sufficiently concentrated, it is essential to choose a suitable distribution $\rho$.
To indicate the subtlety involved, we recall two choices for the case of graphs. Suppose that $G$ consists of edges
$\{u_1,v_1\},\ldots,\{u_m,v_m\}$ and $N_i(x) = |x_{u_i}-x_{v_i}|$ for each $i \in [m]$.
Bencz\'{u}r and Karger \cite{BK96} define $\rho_i$ to be inversely proportional to the largest $k$ such that the edge $\{u_i,v_i\}$ is contained
in a maximal induced $k$-edge-connected subgraph.  Spielman and Srivastava \cite{SS11} define $\rho_i$ as proportional to the effective resistance
across the edge $\{u_i,v_i\}$ in $G$.

\smallskip

Denote the unit ball $B_N \seteq \{ x \in \R^n : N(x) \leq 1 \}$.
We take $\rho_i$ proportional to the average of $N_i(x)$ over the uniform measure on $B_N$:
\begin{equation}\label{eq:nu-def}
   \rho_i \seteq \frac{\int_{B_N} N_i(x)\,dx}{\int_{B_N} N(x)\,dx}\,.
\end{equation}
To motivate this choice of $\rho=(\rho_1,\ldots,\rho_m)$, let us now explain the general framework for
analyzing sparsification by i.i.d. random sampling and chaining.

\paragraph{Symmetrization}
Our goal is to control the maximum deviation
\[\E \max_{x \in B_N} \left|\tilde{N}(x)-\E[\tilde{N}(x)]\right|\,.\]

By a standard symmetrization argument (see \pref{sec:symmetrization}), to bound this quantity by $O(\delta)$, it suffices
to prove that for every {\em fixed} choice of indices $i_1,\ldots,i_M$, we have
\begin{equation}\label{eq:fixed-goal}
   \E_{\e_1,\ldots,\e_M} \frac{1}{M} \sum_{j=1}^M \e_i \frac{N_{i_j}(x)}{\rho_{i_j}} \leq \delta \left(\max_{x \in B_N} \tilde{N}(x)\right)^{1/2},
\end{equation}
where $\e_1,\ldots,\e_M \in \{-1,1\}$ are uniformly random signs.

\paragraph{Chaining and entropy estimates}
If we define $V_x \seteq \frac{1}{M} \left(\e_1 N_{i_1}(x)/\rho_{i_1} + \cdots + \e_M N_{i_M}(x)/\rho_{i_M}\right)$, then $\{ V_x : x \in \R^n \}$
is a subgaussian process, and $\E \max \{ V_x : x \in B_N \}$ can be controlled via standard chaining arguments (see \pref{sec:chaining} for background on subgaussian processes, covering numbers, and chaining upper bounds).
Define the distance
\[
   d(x,y) \seteq \left(\E |V_x-V_y|^2\right)^{1/2} = \frac{1}{M} \left(\sum_{j=1}^M \left(\frac{N_{i_j}(x)-N_{i_j}(y)}{\rho_{i_j}}\right)^2\right)^{1/2}.
\]
and 
let $\cov(B_N, d, r)$ denote the minimum number $K$ such that $B_N$ can be covered by $K$ balls of radius $r$ in the metric $d$.
Then Dudley's entropy bound (\pref{lem:dudley-twice}) asserts that
\begin{equation}\label{eq:dudley}
   \E \max_{x \in B_N} V_x \lesssim \int_0^{\infty} \sqrt{\log \cov(B_N, d, r)}\,dr\,,
\end{equation}
Our goal, then, is to choose sampling probabilities $\rho_1,\ldots,\rho_m$ so as to make the covering numbers $\cov(B_N,d,r)$ suitably small.

In order to get a handle on the distance $d$, let us define
\[
   \cN^\infty(x) \seteq \max_{j \in [M]} \frac{N_{i_j}(x)}{\rho_{i_j}}, \quad \textrm{ and } \quad \kappa \seteq \max \{ \cN^{\infty}(x) : x \in B_N \}\,.
\]
Then we can bound
\begin{align*}
   d(x,y) &\leq M^{-1/2} \sqrt{\cN^{\infty}(x-y)} \left(\frac{1}{M} \sum_{j=1}^M \frac{|N_{i_j}(x)-N_{i_j}(y)|}{\rho_{i_j}} \right)^{1/2} \\
          &\leq M^{-1/2} \sqrt{\cN^{\infty}(x-y)} \left(2 \max_{x \in B_N} \tilde{N}(x)\right)^{1/2}.
\end{align*}
Using this in \eqref{eq:dudley} gives the upper bound
\begin{align}
   \E \max_{x \in B_N} V_x &\lesssim M^{-1/2} \left(\max_{x \in B_N} \tilde{N}(x)\right)^{1/2}\int_0^{\infty} \sqrt{\log \cov(B_N, (\cN^{\infty})^{1/2}, r)}\,dr \nonumber \\
                           &= M^{-1/2} \left(\max_{x \in B_N} \tilde{N}(x)\right)^{1/2}\int_0^{\sqrt{\kappa}} \sqrt{\log \cov(B_N, \cN^{\infty}, r^2)}\,dr\,, \label{eq:dudley2}
\end{align}
where we have used that the last integrand vanishes above $\sqrt{\kappa}$ since $B_N \subseteq \kappa B_{\cN^{\infty}}$.

\paragraph{Dual-Sudakov inequalities}
In order to bound the entropy integral \eqref{eq:dudley2},
let us recall the dual-Sudakov inequality (see \cite{PT85} and \cite[(3.15)]{LedouxTalagrand2011}) which allows one to control covering numbers of the Euclidean ball. Let $B_2^n$ denote the Euclidean ball in $\R^n$. Then for any norm $N$ on $\R^n$, it holds that
\begin{equation}\label{eq:dual-sudakov}
   \sqrt{\log \cov(B_2^n, N, r)} \lesssim \frac{1}{r} \E\left[N(\bm{g})\right]\,,
\end{equation}
where $\bm{g}$ is a standard $n$-dimensional Gaussian.

An adaptation of the Pajor-Talagrand proof of \eqref{eq:dual-sudakov} (see \pref{lem:smooth-sudakov}) allows one to show that for any norms $N$ and $\hat{N}$ on $\R^n$,
\begin{equation}\label{eq:gen-sudakov}
   \log \cov(B_N, \hat{N}, r) \lesssim \frac{1}{r} \E \left[\hat{N} (\bm{Z})\right]\,,
\end{equation}
where $\bm{Z}$ has density proportional to $e^{-N(x)}\,dx$.
A closely related estimate was proved by Milman and Pajor \cite{MP89}; see the remarks after \eqref{eq:dualsudakov}.

In particular, we can apply this with $\hat{N} = \cN^\infty$, yielding
\begin{equation}\label{eq:sudakov-light}
   \log \cov(B_N, \cN^{\infty}, r) \lesssim \frac{1}{r} \E\left[\cN^\infty(\bm{Z})\right] = \frac{1}{r} \E \max_{j \in [M]} \frac{N_{i_j}(\bm{Z})}{\rho_{i_j}}\,.
\end{equation}

At this point, it is quite natural to hope that $N_i(\bm{Z})$ is concentrated around its mean, in which case the choice $\rho_i \propto \E[N_i(\bm{Z})]$ seems appropriate.
We remark that this choice coincides with \eqref{eq:nu-def}. Indeed, the probabilities $\rho_i$ are the same when averaging $N_i$ 
over any density that depends only on $N(x)$.

This is the first point at which we will employ convexity in an essential way. The density $e^{-N(x)}$ is log-concave, and therefore $\bm{Z}$ is a log-concave random variable. By recent progress on the KLS conjecture, one knows that Lipschitz functions of isotropic log-concave vectors concentrate tightly around their mean.

Let $\psi_n$ denote the KLS constant in dimension $n$. In the past few years there has been remarkable progress on bounding $\psi_n$ \cite{Chen21,KL22,JLV22,Klartag23}.
In particular, Klartag and Lehec established that $\psi_n \leq O((\log n)^5)$, and the best current bound is
$\psi_n \leq O(\sqrt{\log n})$ \cite{Klartag23}.

\paragraph{Exponential concentration and the KLS conjecture}
The next lemma expresses a classical connection between exponential concentration and Poincar\'e inequalities \cite{GM83}.
Say that $\f : \R^n \to \R$ is {\em $L$-Lipschitz} if $\|\f(x)-\f(y)\|_2 \leq L \|x-y\|_2$ for all $x,y \in \R^n$.

\begin{theorem}
   \label{thm:kls-concentration}
   There is a constant $c > 0$ such that the following holds.
   Suppose $\bm{X}$ is a random variable on $\R^n$ whose law is isotropic and log-concave.
   Then for every $L$-Lipschitz function $\f : \R^n \to \R$ and $t > 0$,
   \[
      \Pr\left(\left|\f(\bm{X})-\E[\f(\bm{X})]\right| > t\right) \leq 2 e^{-ct/(\psi_n L)}\,.
   \]
\end{theorem}

In \pref{sec:concentration}, we prove the following consequence.

\begin{corollary}\label{cor:norm-concentration}
   There is a constant $c > 0$ such that the following holds.
   Consider a semi-norm $\cN $ on $\R^n$ and a random vector $\bm{Z}$ whose distribution is symmetric and log-concave.
   Then for any $t > 0$,
   \[
      \Pr\left(\left|\vphantom{\tfrac{1}{2}} \cN(\bm{Z} ) -\E[ \cN( \bm{Z} ) ]\right| > t\right) \leq 2 \exp\left(- \frac{c}{\psi_n} \frac{t}{\E[\cN(\bm{Z} )]}\right)\,.
   \]
\end{corollary}

With this in hand, we can immediately use a union bound (see \pref{lem:tailbound}) to obtain
\[
   \E \max_{j \in [M]} \frac{N_{i_j}(\bm{Z})}{\E[N_{i_j}(\bm{Z})]} \lesssim \psi_n \log M\,.
\]
To make $\rho$ a probability measure, we take $\rho_j \seteq \E[N_j(\bm{Z})]/\E[N(\bm{Z})]$ for $j=1,\ldots,m$, and then \eqref{eq:sudakov-light} becomes
\begin{equation}\label{eq:cover2}
   \log \cov(B_N, \cN^{\infty}, r) \lesssim \frac{1}{r} (\psi_n \log M) \E[N(\bm{Z})] = \frac{1}{r} n \psi_n \log (M)\,,
\end{equation}
where the last inequality uses $\E[N(\bm{Z})]=n$, which follows
from a straightforward integration using that the law of $\bm{Z}$ has density proprtional to $e^{-N(x)}$
(\pref{lem:int-Nz}).
Thus we have
\begin{equation}\label{eq:cover3}
   \int_{\sqrt{\kappa}/n^2}^{\sqrt{\kappa}} \sqrt{\log \cov(B_N, \cN^{\infty}, r^2)}\,dr \lesssim \left(n \psi_n \log M\right)^{1/2} \int_{\sqrt{\kappa}/n^2}^{\sqrt{\kappa}} \frac{1}{r}\,dr
   \lesssim \left(n \psi_n \log M\right)^{1/2} \log n\,.
\end{equation}

Standard volume arguments in $\R^n$ (\pref{lem:norm-cover}) allow us to control the rest of the integral:
\begin{equation*}
   \int_{0}^{1/n^2} \sqrt{\log \cov(B_{\cN^{\infty}},\cN^{\infty},r^2)}\,dr \lesssim 1\,,
\end{equation*}
and therefore
\begin{equation*}\label{eq:piece1}
   \int_{0}^{\sqrt{\kappa}/n^2} \sqrt{\log \cov(B_{N},\cN^{\infty},r^2)}\,dr \leq
   \sqrt{\kappa} \int_{0}^{1/n^2} \sqrt{\log \cov(B_{\cN^{\infty}},\cN^{\infty}, r^2)}\,dr \lesssim \sqrt{\kappa}\,.
\end{equation*}
Plugging this and \eqref{eq:cover3} into \eqref{eq:dudley2} gives
\begin{align*}
   \E \max_{x \in B_N} V_x &\lesssim M^{-1/2} \left(\max_{x \in B_N} \tilde{N}(x)\right)^{1/2} \left(\sqrt{\kappa} + \left(n \psi_n \log M \right)^{1/2} \log n\right)\,.
\end{align*}
Finally, observe that \eqref{eq:cover2} gives the bound $\kappa \lesssim n \psi_n \log (M)$, resulting in
\[
   \E \max_{x \in B_N} V_x \lesssim \left(\frac{n \psi_n \log(M) (\log n)^2}{M}\right)^{1/2} \left(\max_{x \in B_N} \tilde{N}(x)\right)^{1/2}.
\]
Choosing $M \asymp \delta^{-2} n (\log n)^2 \psi_n \log(n/\delta)$ yields our desired goal \eqref{eq:fixed-goal}.

\paragraph{Modifications for sums of $p$th powers}
In order to apply these methods to sums of $p$th powers $N(x)^p = N_1(x)^p + \cdots + N_m(x)^p$ for $p > 1$,
we use the natural analog of \eqref{eq:nu-def}:
\begin{equation}\label{eq:l2-probs}
   \rho_i \seteq \frac{\int_{B_N} N_i(x)^p\,dx}{\int_{B_N} N(x)^p\,dx} = \frac{\int_{\R^n} N_i(x)^p\,e^{-N(x)^{\hat{p}}} \,dx}{\int_{\R^n} N(x)^p\,e^{-N(x)^{\hat{p}}} \,dx}\,, \quad \hat{p} = \min(p,2)\,.
\end{equation}
Note that if $p = 2$ and one defines $N_i(x) \seteq |\langle a_i,x\rangle|$, where $a_1,\ldots,a_m \in \R^n$ are the rows
of a full-rank matrix $A \in \R^{m \times n}$, then $\rho_i = \frac{1}{n} \langle a_i, (A^{\top} A)^{-1} a_i\rangle$ are exactly the scaled leverage scores of $A$.

The main hurdle in this setting is that we only establish the analog of \eqref{eq:gen-sudakov} for $p$-uniformly smooth norms:
As shown in \pref{lem:smooth-sudakov},
if $\bm{Z}$ has the law whose density is proportional to $e^{-N(x)^p}$ and $N$ is $p$-uniformly smooth, then
for any norm $\hat{N}$,
\begin{align}
   (\log \cov(B_N, B_{\hat{N}},r))^{1/p} \lesssim \frac{1}{r} \E [\hat{N} (\bm{Z})]\,. \label{eq:dualsudakov}
\end{align}
A closely-related estimate is mentioned in \cite[Eq. (9)]{MP89}.

\paragraph{General norms and block Lewis weights}
To obtain \pref{thm:sos-general} for general norms, we resort to a dimension-dependent version of \eqref{eq:dualsudakov} (see \pref{lem:gen-l2-cover}).
Moreover, we need to augment the sampling probabilities in \eqref{eq:l2-probs} in order to effectively
bound the diameter $\diam(B_N, \cN^{\infty})$. For this, as well as for sums of squares of $\ell_p$ norms (\pref{thm:sos-lp}), in
\pref{sec:sumsmooth} we formulate a generalization of $\ell_p$ Lewis weights, motivated by the construction of weights in \cite{KKTY21b,JLS23,Lee23}.

For a collection of vectors $a_1,\ldots,a_k \in \R^n$, the $\ell_p$ Lewis weights \cite{Lewis78,Lewis79}
result from consideration of the optimization
\begin{equation}\label{eq:lewis-opt}
   \max \{ |\!\det(U)| : \alpha(U) \leq 1 \}\,,
\end{equation}
where $\alpha$ is the norm on linear operators $U : \R^n \to \R^n$ defined by
\[
   \alpha(U) = \left(\sum_{i=1}^k \|Ua_i\|_2^p\right)^{1/p}\,.
\]
In \pref{sec:sumsmooth}, we consider a substantial generalization of this setting where $S_1 \cup \cdots \cup S_m = \{1,\ldots,k\}$
is a partition of the index set.
Given $p_1,\ldots,p_m \geq 2$ and $q \geq 1$, we define the norm
\[
   \alpha(U) \seteq \left(\sum_{j=1}^m \left(\sum_{i \in S_j} \|U a_i\|_2^{p_j}\right)^{q/p_j}\right)^{1/q}\,,
\]
and establish properties of the corresponding optimizer of \eqref{eq:lewis-opt}.

\subsection{Computing the sampling weights via homotopy}
\label{subsec:computing}

In \pref{sec:algorithm}, we present an
algorithm constructing a sparsifier for $N(x) = N_1(x)+\cdots+N_m(x)$ 
that runs in time $n^{O(1)}$ plus the time 
required to do $m (\log n)^{O(1)} + n^{O(1)}$ total evaluations of norms $N_i(y)$ for various $i \in [m]$ and $y \in \R^n$.
It employs a homotopy-type method that has been used for efficient sparsification in multiple settings
(see, e.g., \cite{mp12rowSampArxivV1, KapralovLMMS17, jss18arxiv, AhmadinejadJSS19}).

To compute reasonable overestimates of the sampling weights $\{\rho_i\}$ from \eqref{eq:nu-def}, one approach is to simply sample from the probability measure $\mu$ with density proportional to $e^{-N(x)}$, evaluate the $N_i$ at the sample, and use a scaling of the average evaluation of $N_i$ as the estimate of $\rho_i$.
Sampling from a log-concave distribution, especially those induced by norms, is a well-studied task, and can be done in
$n^{O(1)} \log^{O(1)}(nR/r)$ time if $N$ is $(r, R)$-rounded; see \cite{CV18,JLLV21} and \pref{thm:sample}, \pref{cor:sample}.
If a norm evaluation can be performed in time $\tEval$, this would naively require time $m n^{O(1)} \log^{O(1)}(nR/r) \tEval$,
whereas we would like our algorithms to run in nearly ``input-sparsity time,'' as expressed before.

We first observe that one need only sample from a distribution with density $\propto e^{-\tilde{N}(x)}$ for
some norm $\tilde{N}$ that is $O(1)$-equivalent to $N(x)$. 
Given this fact, for simplicity let us assume that $N$ is a genuine norm that is $(r,R)$-rounded
in the sense that $r \|x\|_2 \leq N(x) \leq R \|x\|_2$ for all $x \in \R^n$.
Define the family of norms $N_t(x) \seteq N(x) + t \|x\|_2$.
For $t = R$, it holds that $N_R(x)$ is $2$-equivalent to the norm $R \|x\|_2$, and sampling
from the distribution with density $\propto e^{-R \|x\|_2}$ is trivial.
Therefore we can construct an $n (\log n)^{O(1)}$-sparse $1/2$-approximation $\tilde{N}_R(x)$
to $N_R(x)$.

Now assuming we have an $n (\log n)^{O(1)}$-sparse $1/2$-approximation $\tilde{N}_t$ to $N_t$ for $r \leq t \leq R$,
we construct a sparsifier for $N_{t/2}(x)$ by sampling from the measure with density $\propto e^{-\tilde{N}_t(x)}$.
This works because $\tilde{N}_t$ is $2$-equivalent to $N_t$, which is $2$-equivalent to $N_{t/2}$.
After $O(\log (R/r))$ iterations, we arrive at sparse norm $\tilde{N}$ that is $O(1)$-equivalent to $N$, and then
by sampling from the distribution with density $\propto e^{-\tilde{N}(x)}$, we are able to construct
a sparse $\e$-approximation to $N$ itself. To handle the case when $N$ is a semi-norm we modify this approach to instead obtain $\tilde{N}(x)$ such that  $\tilde{N}(x) + \epsilon r \norm{x}_2$ is an $\epsilon$-approximation to $N_{\epsilon r}$ and argue that this suffices for $\tilde{N}$ to be an $O(\epsilon)$-approximation of $N$.

\section{Preliminaries}
\label{sec:prelim}

\renewcommand{\S}{\mathbb{S}}
\newcommand{\nnz}{\mathsf{nnz}}
\newcommand{\ma}{\mathbf{A}}
\newcommand{\mb}{\mathbf{B}}
\newcommand{\mi}{\mathbf{I}}
\renewcommand{\O}{\widetilde{O}}
\newcommand{\N}{\mathcal{N}}
\newcommand{\dom}{\mathrm{dom}}

Let us denote $[n] \seteq \{1,2,\dots,n\}$. All logarithms are taken with base $e$ unless
otherwise indicated. 
We use the
notation $a \lesssim b$ if there exists a universal constant $C > 0$ such that $a
\leq C b$, and the notation $a \asymp b$ for the conjunction of $a \lesssim b$ and $b
\lesssim a$. 

\paragraph{Norms vs. semi-norms}
Note that if $N$ is a semi-norm on $\R^n$, then 
$\ker(N) \seteq \{x \in \R^n: N(x) = 0\}$ is a subspace of $\R^n$ and $N$ is a genuine norm on $\ker(N)^{\perp}$. 
Thus, typically, no difficulty is presented in working with semi-norms.
For instance,
one can define the dual semi-norm $N^*(x) \seteq \sup \{ \langle x,y\rangle:y \in \R^n, N(y) \leq 1 \}$, or equivalently
as the norm on $\ker(N)^{\perp}$ that is dual to $N$.
And if $N$ and $\hat{N}$ are $K$-equivalent semi-norms, then $\ker(N)=\ker(\hat{N})$.
In mathematical statements, we use the term ``semi-norm,'' while in informal remarks
and discussions, we may interchange the two terms.

\subsection{Covering numbers, chaining, and subgaussian processes}
\label{sec:chaining}

Consider a metric space $(T,d)$.
A random process $\{V_x : x \in T\}$ is said to be {\em subgaussian with respect to $d$}
if there is a number $\alpha > 0$ such that
\begin{equation}\label{eq:subgaussian-tail}
   \Pr\left(|V_x-V_y| > t\right) \leq \exp\left(\frac{-t^2}{\alpha^2\,d(x,y)^2}\right),\qquad t > 0\,.
\end{equation}
Say that $\{V_x : x \in T\}$ is {\em centered} if $\E[V_x]=0$ for all $x \in T$.

Given a metric space $(T, d)$, define the ball $B(x,r) \seteq \{ y \in T : d(x,y) \leq r \}$.
\begin{definition}[Covering and entropy numbers]
\label{def:cov}
For a number $r > 0$, we define the {\em covering number} $\cov(T,d,r)$ as the
smallest number of balls $\{ B(x_i,r) : i \geq 1 \}$ required to cover $T$. Define the {\em entropy numbers} $e_h(T,d) \seteq \inf \{ r > 0 : \cov(T,d,r) \leq 2^{2^h} \}$ for $h \geq 0$.
\end{definition}

If $T \subseteq \R^n$ and $N$ is a semi-norm on $\R^n$, it induces a natural distance $d(x,y) \seteq N(x-y)$ on $T$. In this case we use the notation $\cov(T,N,r)$ and $e_h(T,N)$ to denote the associated covering and entropy numbers respectively.
We additionally write $B_{N} \seteq \{ x \in \R^n : N(x) \leq 1 \}$ for the unit ball of $N$.

\paragraph{The generic chaining functional}
Recall Talagrand's generic chaining functional \cite[Def. 2.2.19]{TalagrandBook2014}:
\begin{equation}\label{eq:gamma2}
   \gamma_2(T,d) \seteq \inf_{\{\cA_h\}} \sup_{x \in T} \sum_{h=0}^{\infty} 2^{h/2} \diam(\cA_h(x),d)\,,
\end{equation}
where the infimum runs over all sequences $\{\cA_h : h \geq 0\}$ of partitions of $T$ satisfying $|\cA_h| \leq 2^{2^h}$
for each $h \geq 0$.
Note that we use the notation $\cA_h(x)$ for the unique set of $\cA_h$ that contains $x$.
The next theorem constitutes the generic chaining upper bound; see \cite[Thm 2.2.18]{TalagrandBook2014}.

\begin{theorem}\label{thm:chaining}
   If $\{V_x : x \in T\}$ is a centered subgaussian process satisfying \eqref{eq:subgaussian-tail}
   with respect to distance $d$, then
\begin{equation}\label{eq:chaining}
   \E \sup_{x,y \in T} |V_x-V_y| \lesssim \alpha \gamma_2(T,d)\,.
\end{equation}
\end{theorem}

A classical way of controlling $\gamma_2(T,d)$ is given by Dudley's entropy bound (see, e.g., \cite[Prop 2.2.10]{TalagrandBook2014}).
The follow two upper bounds are equivalent up to universal constant factors.

\begin{lemma}[Dudley]
   \label{lem:dudley-twice}
   For any metric space $(T,d)$, it holds that
   \begin{align}
      \gamma_2(T,d) &\lesssim \sum_{h \geq 0} 2^{h/2} e_h(T,d) \label{eq:dudley-v1} \\
      \gamma_2(T,d) &\lesssim \int_0^{\infty} \sqrt{\log \cov(T,d,r)}\,dr\,. \label{eq:dudley-v2}
   \end{align}
\end{lemma}

The next lemma follows from a straightforward volume argument.

\begin{lemma}\label{lem:norm-cover}
   If $N$ is a semi-norm on $\R^n$, then for any $\e > 0$ and $h \geq 0$,
   \begin{align*}
      \cov(B_N,N,\e) &\leq \left(\frac{2}{\e}\right)^n 
      \ \text{ and } \ 
      e_h(B_N,N) \leq 2 \cdot 2^{-2^h/n}\,.
   \end{align*}
\end{lemma}

To show that our sampling algorithms succeed with high probability, as opposed to only in expectation, we use the following refinement of \pref{thm:chaining}.

\begin{theorem}[{\cite[Thm 2.2.27]{TalagrandBook2014}}]
   \label{thm:fernique-talagrand}
   Suppose $\{V_x : x \in T\}$ is a centered subgaussian process with respect to the distance $d$.
   Then for some constants $c > 0, C > 1$ and any $\lambda > 0$,
   \[
      \Pr\left(\sup_{x,y \in T} |V_x-V_y| > C \left(\vphantom{\hat{h}}\gamma_2(T,d) + \lambda \diam(T,d)\right)\right) \lesssim \exp\left(-c \lambda^2\right)\,.
   \]
   In particular, if $Z = \sup_{x,y\in T} |V_x-V_y|$, then for any $\lambda > 0$,
   \begin{equation}\label{eq:mgf}
      \log \E[e^{\lambda Z}] \lesssim \lambda^2 \diam(T,d)^2 + \lambda \gamma_2(T,d)\,.
   \end{equation}
\end{theorem}

\subsection{Sparsification via subgaussian processes}
\label{sec:symmetrization}

Consider $\f_1,\f_2,\ldots,\f_m : \R^n \to \R$, and define
\[
   F(x) \seteq \sum_{j \in [m]} \f_j(x)\,.
\]
Given a probability vector $\rho \in \R_+^m$,
and an integer $M \geq 1$ and $\nu=(\nu_1,\ldots,\nu_M) \in [m]^M$,
define the distance
\begin{equation}
\label{eq:d_rho_nu}
   d_{\rho,\nu}(x,y) \seteq \left(\sum_{j \in [M]} \left(\frac{\f_{\nu_j}(x)-\f_{\nu_j}(y)}{\rho_{\nu_j}\cdot M}\right)^2\right)^{1/2}.
\end{equation}
and the function $\tilde{F}_{\rho,\nu} : \R^n \to \R$ 
\begin{align*}
   \tilde{F}_{\rho,\nu}(x) &\seteq \frac{1}{M} \sum_{j \in [M]} \frac{\f_{\nu_j}(x)}{\rho_{\nu_j}}\,.
\end{align*}

\newcommand{\Comega}[1]{\left\|#1\right\|_{C(\Omega)}}
\newcommand{\Fomega}{\Comega{F}}
\newcommand{\Ftomega}{\Comega{\tilde{F}_{\rho,\nu}}}
\newcommand{\Ftomegabold}{\Comega{\tilde{F}_{\rho,\bm{\nu}}}}

The next lemma employs a variant of a standard symmetrization argument to control $\E \max_{x \in \Omega} |F(x)-\tilde{F}_{\rho,\bm{\nu}}(x)|$ using an associated subgaussian process (see, for example, \cite[Lem 9.1.11]{TalagrandBook2014}). We also prove a version with a tail bound to show that our algorithms succeed with high probability.

For a subset $\Omega \subseteq \R^n$, we use the notation $\Fomega \seteq \sup_{x \in \Omega} |F(x)|$.
Note that in every application of the next lemma in the present paper, we take $\Omega = \{ x \in \R^n : F(x) \leq 1 \}$.

\begin{lemma}
   \label{lem:sparsification-meta}
   Consider $M \geq 1$, a subset $\Omega \subseteq \R^n$, and a probability vector $\rho \in \R^m_+$.
   Assume that
   \begin{equation}\label{eq:haszero}
      \exists x_0 \in \Omega \quad \textrm{s.t.} \quad \f_1(x_0)=\cdots=\f_m(x_0)=0\,.
   \end{equation}
   Suppose, further, that for some $0 < \delta \leq 1$, and every $\nu \in [m]^M$, it holds that
   \begin{equation}\label{eq:gamma-assumption}
      \gamma_2(\Omega, d_{\rho,\nu}) \leq \delta \left(\Fomega \Ftomega\right)^{1/2}\,.
   \end{equation}
   If $\bm{\nu}_1,\ldots,\bm{\nu}_M$ are sampled independently from $\rho$, then
   \begin{equation}\label{eq:sm-expectation}
      \E \max_{x \in \Omega} \left|F(x)-\tilde{F}_{\rho,\bm{\nu}}(x)\right| \lesssim \E\left[\gamma_2(\Omega,d_{\rho,\bm{\nu}})\right] \leq 8 \delta \Comega{F}.
   \end{equation}
   If it also holds that, for all $\nu \in [m]^M$,
   \begin{equation}\label{eq:diam-assumption}
      \diam(\Omega, d_{\rho,\nu}) \leq \bdelta \left(\Fomega \Ftomega\right)^{1/2}\,,
   \end{equation}
   then there is a universal constant $K > 0$ such that for all $0 \leq t \leq \frac{1}{2K\bdelta}$,
   \begin{equation}\label{eq:max-tail}
      \Pr \left(\max_{x \in \Omega} \left|F(x) - \tilde{F}_{\rho,\bm{\nu}}(x)\right| > K(\delta + t \bdelta) \Fomega \right) \leq e^{-Kt^2/4}.
   \end{equation}
\end{lemma}
\begin{proof}
Note that $\E[\tilde{F}_{\rho,\bm{\nu}}(x)]=F(x)$ for every $x \in \R^n$.
Thus for any convex function $\psi : \R_+ \to \R_+$,
\begin{equation}\label{eq:start}
   \E_{\bm{\nu}} \psi\left(\max_{x \in \Omega} \left|F(x)-\tilde{F}_{\rho,\bm{\nu}}(x)\right|\right) \leq
   \E_{\bm{\nu},\tilde{\bm{\nu}}} \psi\left(\max_{x \in \Omega} \left|\tilde{F}_{\rho,\bm{\nu}}(x)-\tilde{F}_{\rho,\tilde{\bm{\nu}}}(x)\right|\right)\,,
\end{equation}
where $\tilde{\bm{\nu}}$ is an independent copy of $\bm{\nu}$.
The argument to $\psi$ on the right-hand side can be written as
\[
   \max_{x \in \Omega} \left|\frac{1}{M} \sum_{j \in [M]} \left(\frac{\f_{\bm{\nu}_j}(x)}{\rho_{\bm{\nu}_j}} -
   \frac{\f_{\tilde{\bm{\nu}}_j}(x)}{\rho_{\tilde{\bm{\nu}_j}}}\right)\right|.
\]
Since the distribution of $\f_{\bm{\nu}_j}(x)/\rho_{\bm{\nu}_j} - \f_{\tilde{\bm{\nu}}_j}(x)/\rho_{\tilde{\bm{\nu}}_j}$ is symmetric, we have
\[
   \sum_{j \in [M]} \frac{\f_{\bm{\nu}_j}(x)}{\rho_{\bm{\nu}_j}} - \sum_{j \in [M]} \frac{\f_{\tilde{\bm{\nu}}_j}(x)}{\rho_{\tilde{\bm{\nu}}_j}}
   \ \stackrel{\mathrm{law}}{=}\ 
\sum_{j \in [M]} \e_j \cdot \left(\frac{\f_{\bm{\nu}_j}(x)}{\rho_{\bm{\nu}_j}} - \frac{\f_{\tilde{\bm{\nu}}_j}(x)}{\rho_{\tilde{\bm{\nu}}_j}}\right)
\]
for any choice of signs $\e_1,\ldots,\e_M \in \{-1,1\}$. This yields
the stochastic domination
\begin{align}
   \max_{x \in \Omega} \left|\frac{1}{M} \sum_{j = 1}^{M} \frac{\f_{\bm{\nu}_j}(x)}{\rho_{\bm{\nu}_j}}- \frac{1}{M} \sum_{j = 1}^{M} \frac{\f_{\tilde{\bm{\nu}}_j}(x)}{\rho_{\tilde{\bm{\nu}}_j}}\right|
   &\preceq
   \max_{x \in \Omega} \left|\frac{1}{M} \sum_{j = 1}^{M} \e_j \frac{\f_{\bm{\nu}_j}(x)}{\rho_{\bm{\nu}_j}}\right| + 
   \max_{x \in \Omega} \left|\frac{1}{M} \sum_{j = 1}^{M} \e_j \frac{\f_{\tilde{\bm{\nu}}_j}(x)}{\rho_{\tilde{\bm{\nu}}_j}}\right|. \label{eq:sym}
\end{align}
Note that if we choose $\e_1,\ldots,\e_M \in \{-1,1\}$ to be uniformly random, then
the quantity in the absolute value is a centered subgaussian process with respect to the distance $d_{\rho,\bm{\nu}}$ on $\Omega$,
so we are in position to apply \pref{thm:fernique-talagrand}.

Define $\cS \seteq \E \max_{x \in \Omega} |F(x)-\tilde{F}_{\rho,\bm{\nu}}(x)|$ and apply \eqref{eq:start} with $\psi(x)=x$ and \eqref{eq:sym} to obtain 
\begin{align}
   \cS &\leq 2 \E_{\bm{\nu}}\E_{\e}  \max_{x \in \Omega} \left|\frac{1}{M} \sum_{j = 1}^{M} \e_j \frac{\f_{\bm{\nu}_j}(x)}{\rho_{\bm{\nu}_j}}\right| \nonumber \\
       &= 2 \E_{\bm{\nu}}\E_{\e}  \max_{x \in \Omega} \left|\frac{1}{M} \sum_{j = 1}^{M} \e_j \frac{\f_{\bm{\nu}_j}(x)-\f_{\bm{\nu}_j}(x_0)}{\rho_{\bm{\nu}_j}}\right|
   \lesssim  \E_{\bm{\nu}} [\gamma_2(\Omega, d_{\rho,\bm{\nu}})]\,,\label{eq:Sbnd1}
\end{align}
where the equality follows from \eqref{eq:haszero}, and
the second inequality is an application of \pref{thm:chaining}.

Now use \eqref{eq:gamma-assumption} and concavity of the square root to bound
\begin{equation}\label{eq:gbnd}
   \E [\gamma_2(\Omega, d_{\rho,\bm{\nu}})] \leq \delta \left(\Fomega \E \Ftomegabold\right)^{1/2}
\end{equation}
The triangle inequality gives
\[
   \Ftomega \leq \Fomega + \max_{x \in \Omega} |F(x)-\tilde{F}_{\rho,\bm{\nu}}(x)|\,.
\]
In conjunction with \eqref{eq:Sbnd1} and \eqref{eq:gbnd}, this yields the consequence
\[
   \cS \lesssim \delta \Fomega^{1/2} \left(\Fomega+\cS\right)^{1/2} = \delta \Fomega \left(1+\Fomega^{-1} \cS\right)^{1/2}.
\]
Since $\delta \leq 1$, this confirms \eqref{eq:sm-expectation}.
\smallskip

Let us now verify \eqref{eq:max-tail}. 
Fix $\lambda > 0$ and define $\bm{Z} \seteq \max_{x \in \Omega} |F(x)-\tilde{F}_{\rho,\bm{\nu}}(x)|$.
Applying \eqref{eq:start} with $\psi(x)=e^{\lambda x}$ and \eqref{eq:sym}, yields
\begin{align*}
   \E[e^{\lambda \bm{Z}}] =
   \E_{\bm{\nu}} \exp\left(\lambda \max_{x \in \Omega} \left|F(x)-\tilde{F}_{\rho,\bm{\nu}}(x)\right|\right) &\leq
   \E_{\bm{\nu}} \E_{\e_1,\ldots,\e_M} \exp\left(2 \lambda \max_{x \in \Omega}
   \left|\frac{1}{M} \sum_{j \in [M]} \e_j \frac{\f_{\bm{\nu}_j}(x)-\f_{\bm{\nu}_j}(x_0)}{\rho_{\bm{\nu}_j}}\right|\right)\\
   &\leq
   \E_{\bm{\nu}} \exp\left(C \left(\lambda^2 \diam(\Omega, d_{\rho,\bm{\nu}})^2 + \lambda \gamma_2(\Omega, d_{\rho,\bm{\nu}}) \right)\right)\,,
\end{align*}
where the last inequality is an invocation of \eqref{eq:mgf}.
Using \eqref{eq:gamma-assumption} and \eqref{eq:diam-assumption}, the latter quantity is bounded by
\begin{align*}
   \E_{\bm{\nu}} \exp&\left(C \left(\lambda^2  \bdelta^2 \Fomega \Ftomega + \lambda \delta \left(\Fomega \Ftomega\right)^{1/2}\right)\right)\\ 
                &\leq
                \E \exp\left(C \lambda^2 \bdelta^2 \Fomega^2 (1+ \Fomega^{-1} \bm{Z}) + C \lambda \delta \Fomega \left(1 + \Fomega^{-1} \bm{Z}\right)^{1/2}\right)\,.
\end{align*}
Observe that for any $\alpha > 0$ and $z \geq 0$, we have $(1+z)^{1/2} \leq (1+\alpha)^{1/2} + \alpha^{-1/2} z$.
Choose $\alpha \seteq (4 C \delta)^2$ so that
\[
   \E[e^{\lambda \bm{Z}}] \leq \exp\left(C \lambda^2 \bdelta^2 \Fomega^2 + C\lambda \delta \Fomega (1+(4C\delta)^2)^{1/2}\right) \E \exp\left(\left(C \Fomega\lambda^2 \bdelta^2 + \lambda/4\right)\bm{Z}\right)\,.
\]
Let us now assume that $t \leq (2 C \bdelta)^{-1}$ and
choose $\lambda \seteq t/(2\bdelta \Fomega)$. In this case, $C \Fomega \lambda^2 \bdelta^2 \le \lambda/4$, and therefore
the last factor on the right-hand side is at most $\E e^{\lambda \bm{Z}/2} \leq (\E e^{\lambda \bm{Z}})^{1/2}$.
So we arrive at the bound
\begin{align*}
   \E[e^{\lambda \bm{Z}}] 
   &\leq
   \exp\left(C \lambda^2 \bdelta^2 \Fomega^2 + C\lambda \delta \Fomega (1+(4C\delta)^2)^{1/2}\right)
    (\E e^{\lambda \bm{Z}})^{1/2}\,.
\end{align*}
Taking logs and using $\delta \leq 1$ gives
\[
   \log \E[e^{\lambda \bm{Z}}] \leq K(\lambda^2 \bdelta^2 \Fomega^2 + \lambda \delta \Fomega)
\]
for some universal constant $K > 0$.
Let us finally observe the standard consequence of Markov's inequality,
\begin{align*}
   \log \Pr\left(\bm{Z} > K(\delta + t \bdelta) \Fomega\right) &\leq \log \E[e^{\lambda \bm{Z}}] - \lambda K(\delta + t \bdelta)\Fomega \\
                                                               &\leq K(\lambda^2\bdelta^2 \Fomega^2 - t\lambda\bdelta \Fomega)  = -Kt^2/4\,,
\end{align*}
completing the proof.
\end{proof}

\subsection{Concentration for Lipschitz functionals}
\label{sec:concentration}

We use the following standard concentration result for $n$-dimensional Gaussians (see, e.g., \cite[(1.4)]{LedouxTalagrand2011}).

\begin{theorem}
   \label{thm:gaussian-concentration}
   Let $\bm{g}$ be a standard $n$-dimensional Gaussian. Then for every $L$-Lipschitz function $\f: \R^n \to \R$ and $t > 0$,
   \[
      \Pr\left(\left|\f(\bm{g}) - \E \f(\bm{g})\right| > t\right) \leq \exp\left(-t^2/(2L^2)\right).
   \]
\end{theorem}

We also use a classical moment inequality (see, e.g., \cite[Rem. 5]{LW08} for computation of the constant).

\begin{theorem}[\cite{BPF63}]
   \label{thm:lc-moments}
   Suppose $\bm{X}$ is a real, symmetric, log-concave random variable. Then for every $p \geq q > 0$,
   it holds that
   \[
      (\E |\bm{X}|^p)^{1/p} \leq \frac{p}{q} (\E |\bm{X}|^q)^{1/q}\,.
   \]
\end{theorem}

Together with \pref{thm:kls-concentration}, this yields \pref{cor:norm-concentration}, as we now show.

\begin{proof}[Proof of \pref{cor:norm-concentration}]
   Define the covariance matrix $A \seteq \E[\bm{Z} \bm{Z}^{\top}]$ and let $\bm{X} \seteq A^{-1/2} \bm{Z}$.
   Then the law of $\bm{X}$ is log-concave and isotropic by construction.
   Thus \pref{thm:kls-concentration} gives the desired result once we confirm the Lipschitz bound
   \begin{equation}\label{eq:goal}
       \cN (A^{1/2} x) \leq 2 \E[\cN(\bm{Z}) ]\cdot \|x\|_2\,.
   \end{equation}

   To this end, let $\cN^*$ denote the dual semi-norm to $\cN$ and write
   \begin{align*}
   \cN (A^{1/2} x) &= \sup_{\cN^*(w)\leq 1} \langle w, A^{1/2} x\rangle \\
                  &= \sup_{\cN^*(w)\leq 1} \langle A^{1/2} w, x \rangle 
                   \leq \|x\|_2 \sup_{\cN^*(w)\leq 1} \|A^{1/2} w\|_2\,.
   \end{align*}
   Then we have
   \[
      \|A^{1/2} w\|_2 = \langle w, A w\rangle^{1/2} = \left(\E [\langle w,\bm{Z}\rangle^2]\right)^{1/2} \leq 2 \E[|\langle w,\bm{Z}\rangle|] \leq 2 \cN^*(w) \E[\cN(\bm{Z}) ]\,,
   \]
   where the penultimate inequality follows from \pref{thm:lc-moments}, since $\langle y,\bm{Z}\rangle$ is a symmetric log-concave random variable.
\end{proof}

\section{Sparsification for uniformly smooth norms}
\label{sec:sum}

\newcommand{\hatN}{\hat{N}}
\newcommand{\hatcN}{\hat{\cN}}
\newcommand{\bnorm}{\mathscr{N}}

This section studies sparsification of uniformly smooth norms, and shows \pref{thm:l1-sparsification} and \pref{thm:l2-sparsification}.
We start by establishing a variant of the dual Sudakov lemma that will allow us to bound certain covering numbers.

\subsection{Dual Sudakov lemmas for smooth norms}

\newcommand{\sudnorm}{N}
\newcommand{\hatsudnorm}{\widehat{N}}
\newcommand{\convexbody}{W}
\newcommand{\smoothconst}{S}

\begin{lemma}[Shift lemma]
   \label{lem:shift}
   Suppose $\sudnorm$ is a norm on $\R^n$ that is $p$-uniformly smooth with constant $\smoothconst_p$. Define the probability measure $\mu$ on $\R^n$ by
   \[
      d\mu(x) \propto \exp(-\sudnorm(x)^p)\,dx\,.
   \]
   Then for any symmetric convex body ${\convexbody}$ and $z\in \R^n$,
   \begin{equation}\label{eq:shift-measure}
      \mu({\convexbody}+z) \geq \exp(- \smoothconst_p^p \sudnorm (z)^p)\,\mu({\convexbody})\,.
   \end{equation}
\end{lemma}

\begin{proof}
   For any $z \in \R^n$, it holds that
   \[
      \mu({\convexbody}+z) =  \frac{\int_{\convexbody} \exp(-\sudnorm(x+z)^p)\,dx}{\int_{\convexbody} \exp(-\sudnorm(x)^p)\,dx} \mu({\convexbody})\,.
   \]
   Now we bound
   \begin{align*}
      \int_{\convexbody} \exp(-\sudnorm(x+z)^p)\,dx
                              &=  \int_{\convexbody} \E_{\sigma \in \{-1,1\}} \exp(-\sudnorm(\sigma x+z)^p)\,dx \\
                        &\geq  \int_{\convexbody} \exp\left(-\E_{\sigma \in \{-1,1\}} \sudnorm(\sigma x+z)^p\right)\,dx  \\
                        &\geq   \int_{\convexbody} \exp\left(-(\sudnorm(x)^p + \smoothconst_p^p \sudnorm(z)^p)\right)\,dx  \\
                        &= \exp(-\smoothconst_p^p \sudnorm(z)^p) \int_{\convexbody} \exp(-\sudnorm(x)^p)\,dx\,,
   \end{align*}
   where the equality uses symmetry of ${\convexbody}$, the first inequality uses convexity of $\exp(x)$, and the second inequality
   uses $p$-uniform smoothness of $\sudnorm$ (recall \eqref{eq:p-smooth}).
\end{proof}

\begin{lemma}\label{lem:smooth-sudakov}
   Let $\sudnorm$ and $\hatsudnorm$ be semi-norms on $\R^n$ such that $\ker(\sudnorm) \subseteq \ker(\hatsudnorm)$.
   Suppose that $\sudnorm$ is $p$-uniformly smooth with constant $\smoothconst_p$, 
   and define the probability measure $\mu$ on $\ker(N)^{\perp}$ so that 
   \[
      d\mu(x) \propto \exp(-\sudnorm(x)^p)\,dx\,.
   \]
   Then for any $\e > 0$,
   \[
      \left(\log \left(\cov(B_{\sudnorm}, \hatsudnorm, \e)/2\right)\right)^{1/p} \leq 2 \frac{\smoothconst_p}{\e} \int \hatsudnorm(x)\,d\mu(x)\,.
   \]
\end{lemma}

\begin{proof}
   By scaling $\hatsudnorm$, we may assume that $\e = 1$. 
Suppose now that $x_1,\ldots,x_M \in B_{\sudnorm}$ and $x_1+B_{\hatsudnorm},\ldots,x_M+B_{\hatsudnorm}$ are pairwise disjoint.
To establish an upper bound on $M$, let $\lambda > 0$ be a number
we will choose later and write
\begin{align*}
   1 \geq \mu\left(\bigcup_{j \in [M]} \lambda (x_j + B_{\hatsudnorm})\right)  &= \sum_{j \in [M]} \mu\left(\lambda x_j + \lambda B_{\hatsudnorm}\right) \\
     &\stackrel{\mathclap{\eqref{eq:shift-measure}}}{\geq}\ \sum_{j \in [M]} e^{-\lambda^p \smoothconst_p^p \sudnorm(x_j)^p} \mu(\lambda B_{\hatsudnorm})
     \geq M e^{-\smoothconst_p^p \lambda^p} \mu(\lambda B_{\hatsudnorm})\,,
\end{align*}
where \eqref{eq:shift-measure} used Lemma~\ref{lem:shift} and the last inequality used $x_1,\ldots,x_M \in B_\sudnorm$.

Now choose $\lambda \seteq 2 \int \hatsudnorm(x)\, d\mu(x)$
so that Markov's inequality gives
\[
   \mu(\lambda B_{\hatsudnorm}) = \mu\left(\{x :  \hatsudnorm(x) \leq \lambda \} \right) \geq 1/2\,.
\]
Combining with the preceding inequality yields the upper bound
\[
   \left(\log (M/2)\right)^{1/p} \leq \smoothconst_p \lambda\,.\qedhere
\]
\end{proof}

\subsection{Entropy estimates}
\label{sec:entropy}

Here we establish our primary entropy estimate.

\begin{definition}
   \label{def:entropy-objects}
   Consider $p \geq 1$ and denote $\hat{p} \seteq \min(p,2)$.
   Let $\bnorm$ be a semi-norm on $\R^n$ that 
   is $\hat{p}$-uniformly smooth with constant $\smoothconst$.
   Denote by $\mu$ the probability measure on $\ker(\bnorm)^{\perp}$ with $d\mu(x) \propto e^{-\bnorm(x)^{\hat{p}}}\,dx$. 
   Consider any semi-norms $\cN_1,\ldots,\cN_M$ on $\R^n$ with $\ker(\cN_1),\ldots,\ker(\cN_M) \supseteq \ker(\bnorm)$, and define
   \begin{align*}
      \mathsf{w}_j &\seteq \int \cN_j(x)\,d\mu(x),\quad j \in [M] \\
      \mathsf{w}_{\infty} &\seteq \max\left(\mathsf{w}_1,\ldots,\mathsf{w}_M\right) \\
      d(x,y) &\seteq \left(\sum_{j=1}^M (\cN_j(x)^p-\cN_j(y)^p)^2\right)^{1/2} \\
      \cN^{\infty}(x) &\seteq \max_{j \in [M]} \cN_j(x) \\
      \Lambda_{\Omega} &\seteq \sup_{x \in \Omega} \sum_{j=1}^M \cN_j(x)^p\,,
   \end{align*}
   for some $\Omega \subseteq B_{\bnorm}$.
\end{definition}
We begin with some preliminary bounds valid for all $p \geq 1$.
The next lemma will be useful in controlling the diameter of $(\Omega,d)$.

\begin{lemma}\label{lem:lp-diam}
   For any semi-norm $\hatN$ on $\R^n$ with $\ker(\bnorm) \subseteq \ker(\hatN)$,
   \[
      \max_{x \in B_{\bnorm}} \hatN(x) \leq 12 \smoothconst \int \hatN(x) \,d\mu(x)\,.
   \]
\end{lemma}

\begin{proof}
   Define $m \seteq \int \hatN(x) \,d\mu(x)$. Markov's inequality gives $\mu(\lambda B_{\hatN}) \geq 3/4$, where $\lambda \seteq 4m$. Now \pref{lem:shift} 
   (applied with $p=\hat{p}$) gives
\[
   \mu(\lambda B_{\hatN} + y) \geq \exp \left( - \smoothconst^{\hat{p}} \bnorm(y)^{\hat{p}} \right) \mu(\lambda B_{\hatN}) \geq \frac{3}{4} \exp \left( - \smoothconst^{\hat{p}} \bnorm(y)^{\hat{p}} \right)\,.
\]
If $\bnorm(y) \leq 3^{-1/\hat{p}}/\smoothconst$, this implies $\mu(\lambda B_{\hatN} + y), \mu(\lambda B_{\hatN} - y) > 1/2$. Thus
$(\lambda B_{\hatN} + y) \cap (\lambda B_{\hatN} - y) \neq \emptyset$, and
therefore some $z \in \lambda B_{\hatN}$ satisfies $z+y,z-y \in \lambda B_{\hatN}$.
By convexity, we have $y \in \lambda B_{\hatN}$ as well, i.e., $\hatN(y) \leq \lambda$. Since this holds for any $y$ satisfying $\bnorm(y) \leq 3^{-1/\hat{p}}/\smoothconst$, the claim follows.
\end{proof}

Applying the preceding lemma with $\hat{N}=\cN_j$ for each $j=1,\ldots,M$ gives the following.

\begin{corollary}\label{cor:pre-diameter-bound}
   It holds that
   \[
      \diam(B_{\bnorm}, \cN^{\infty})  \lesssim \smoothconst \max_{j \in [M]} \int \cN_j(x)\,d\mu(x) = \smoothconst \mathsf{w}_{\infty}\,.
   \]
\end{corollary}

We recall the following basic maximal inequality.

\begin{fact}\label{fact:maximal}
   If $\bm{X}_1,\ldots,\bm{X}_M$ are nonnegative random variables satisfying
   $\Pr[\bm{X}_j \geq 1 + t] \leq C \exp(-t/\beta)$ for $t > 0, j \in [M]$ and some $C,\beta \geq 1$,
   then $\E[\max_{j \in [M]} \bm{X}_j] \lesssim \beta \left(1+\log (CM)\right)$.
\end{fact}

\begin{proof}
   A union bound gives $\Pr[\textstyle{\max_j} \bm{X}_j \geq 1+t] \leq C M e^{-\beta t}$, therefore for any $\theta > 1$,
   \[
      \E[\textstyle{\max_j \bm{X}_j}] = \int_0^{\infty} \Pr[\textstyle{\max_j \bm{X}_j} \geq t] \leq \theta + C M\int_{\theta}^{\infty} e^{-\beta (t-1)}\,dt
      = \theta + C\beta M e^{\beta-\theta/\beta}\,.
   \]
   Choosing $\theta \seteq \beta (1+2 \log(CM))$ gives $\E[\max_j \bm{X}_j] \lesssim \beta (1+\log(CM))$.
\end{proof}

\begin{lemma}\label{lem:tailbound}
   It holds that
\begin{align*}
   \int \cN^{\infty}(x) \,d\mu(x) &\lesssim \psi_n \log(M)\,\mathsf{w}_{\infty}\,.
\end{align*}
\end{lemma}

\begin{proof}
   Suppose that $\bm{Z}$ has law $\mu$, and define
   $\bm{X}_j \seteq \cN_j(\bm{Z})/\E[\cN_j(\bm{Z})]$ for $j \in [M]$.
   Note that
   \[
      \int \cN^{\infty}(x)\,d\mu(x) = \E[\textstyle{\max_{j}} \cN_j(\bm{Z})] \leq \textstyle{\max_{j}} \E[\cN_j(\bm{Z})] \cdot
      \E [\textstyle{\max_j} \bm{X}_j]\,.
   \]
   \pref{cor:norm-concentration} asserts that $\Pr[\bm{X}_j \geq t + 1] \leq 2 e^{- ct/\psi_n}$, and therefore \pref{fact:maximal} gives
   $\E[\max_j \bm{X}_j] \lesssim \psi_n \log M$, establishing the first claimed inequality.
\end{proof}

For the remainder of this subsection, we restrict ourself to the range $p \in [1,2]$.
We will control $d$ by $\cN^{\infty}$ using the next estimate.

\begin{lemma}
\label{lem:inside0}
For all $x,y \in \Omega$,
\begin{equation}
   d(x,y)^2 \leq 4\Lambda_{\Omega} \left(\cN^{\infty}(x-y)\right)^{p}\,.
\end{equation}
\end{lemma}

\begin{proof}
   Monotonicity of $q$th powers implies that for all $u,v \in \R$ we have $|u+v|^q \leq |u|^q + |v|^q$ for $q \in [0,1]$.
   Applying this with $u=a-b$ and $v=b$ gives $|a^q-b^q| \leq |a-b|^q$.
   Thus for real numbers $a,b \geq 0$ and $p \in [1,2]$, we have
   \[
      |a^p-b^p| = |a^{p/2}-b^{p/2}| |a^{p/2}+b^{p/2}| \leq |a-b|^{p/2} |a^{p/2}+b^{p/2}|\,.
   \]
   Squaring both sides yields
   \[
      |a^p-b^p|^2 \leq 2 |a-b|^{p} (a^p + b^p)\,.
   \]
   Applying this with $a=\cN_j(x),b=\cN_j(y)$ for each $j \in [M]$ we arrive at
   \[
      d(x,y)^2 \leq 2 \sum_{j=1}^M |\cN_j(x)-\cN_j(y)|^p (\cN_j(x)^p+\cN_j(y)^p) \leq 4 \Lambda_{\Omega} \max_{j \in [M]} |\cN_j(x)-\cN_j(y)|^p\,.
   \]
   Finally, note that the triangle inequality gives $|\cN_j(x)-\cN_j(y)| \leq \cN_j(x-y)$ for each $j \in [M]$, completing the proof.
\end{proof}

In conjunction with \pref{cor:pre-diameter-bound}, this yields the following.

\begin{corollary}\label{cor:diameter-bound}
   For $p \in [1,2]$, it holds that $\diam(\Omega, d) \lesssim (\smoothconst \mathsf{w}_{\infty})^{p/2} \sqrt{\Lambda_{\Omega}}$.
\end{corollary}

We now prove our primary entropy estimate.

\begin{lemma}[Entropy bound]\label{lem:lp-chaining}
   It holds that
   \begin{equation}\label{eq:lp-chaining} 
      \gamma_2(\Omega, d) \lesssim \left(\smoothconst \mathsf{w}_{\infty} \psi_n \log M\right)^{p/2} \log(n) \sqrt{\Lambda_{\Omega}}\,.
   \end{equation}
\end{lemma}

\begin{proof}
Since both sides of \eqref{eq:lp-chaining} scale linearly in the $p$th powers $\{\cN_j^p\}$, we may assume that
   \begin{equation*}
      \Lambda_{\Omega} = \max_{x \in \Omega} \sum_{j=1}^M \cN_j(x)^p = 1\,.
   \end{equation*}
Thus \pref{lem:inside0} gives
\begin{equation}\label{eq:inside}
   d(x,y) \leq 2 \cN^\infty(x-y)^{p/2}\,,\qquad x,y \in \Omega\,.
\end{equation}
We first claim that
\begin{align}
   \left(\log \cov(B_{\bnorm}, \cN^{\infty}, \e)\right)^{1/p} &\lesssim \frac{\smoothconst}{\e} \int \cN^{\infty}(x) \,d\mu(x) 
   \lesssim  \frac{\smoothconst}{\e} \psi_n \log(M) \mathsf{w}_{\infty}\label{eq:lp-cov-2} \,.
\end{align}
The first inequality uses \pref{lem:smooth-sudakov} with $N = \bnorm, \hatsudnorm = \cN^\infty$, and the second uses \pref{lem:tailbound}.

Define $Q \seteq \psi_n \log(M) \mathsf{w}_{\infty}$ and then using \eqref{eq:inside} and $\Omega \subseteq B_{\bnorm}$, we have
\[
\sqrt{\log \cov(\Omega, d, \e)} \lesssim \sqrt{\log \cov(B_{\bnorm}, \cN^{\infty}, (\e/\sqrt{2})^{2/p})} \stackrel{\eqref{eq:lp-cov-2}}{\lesssim} \frac{(\smoothconst Q)^{p/2}}{\e}\,,
\]
which immediately yields
\begin{equation}\label{eq:pre-dud1}
   e_h(\Omega, d) \lesssim 2^{-h/2} (\smoothconst Q)^{p/2}\,,\qquad h \geq 0\,.
\end{equation}
Using \eqref{eq:inside} again gives
\begin{eqnarray}
e_h(\Omega, d) \lesssim e_h(B_{\bnorm},\cN^{\infty})^{p/2} &\lesssim& \diam(B_{\bnorm},\cN^{\infty})^{p/2}\ e_h(B_{\cN^{\infty}}, \cN^{\infty})^{p/2} \nonumber
\\
                                                                 &\lesssim& \diam(B_{\bnorm},\cN^{\infty})^{p/2}\ 2^{-2^h p/2n}\,,\label{eq:pre-dud2}
\end{eqnarray}
where the second inequality follows from $B_{\bnorm} \subseteq \diam(B_{\bnorm},\cN^{\infty}) \cdot B_{\cN^\infty}$ and the final inequality is from \pref{lem:norm-cover}.

Then using \eqref{eq:pre-dud1} and \eqref{eq:pre-dud2} in conjunction with the Dudley entropy bound \eqref{eq:dudley-v1} gives
\begin{align*}
   \gamma_2(\Omega, d) &\lesssim \sum_{0 \leq h \leq 4 \log n} 2^{h/2} e_h(\Omega,d) + \diam(B_{\bnorm},\cN^{\infty})^{p/2}\sum_{h > 4 \log n} 2^{h/2} 2^{-2^h p/2n} \nonumber \\
                    &\lesssim (\smoothconst Q)^{p/2} \log n + \diam(B_{\bnorm},\cN^{\infty})^{p/2}\,.
\end{align*}
In conjunction with \pref{cor:pre-diameter-bound}, the proof is complete.
\end{proof}

\subsubsection{Entropy for $p \geq 2$}

We will continue working under the definition \pref{def:entropy-objects}, but now restrict ourselves
to the regime $p \geq 2$, where $\hat{p}=2$.
We additionally define the quantity
\begin{equation}\label{eq:tlam}
   \tilde{\Lambda}_{\Omega} \seteq \sup_{x \in \Omega} \sum_{j=1}^M \cN_j(x)^{2(p-1)}\,.
\end{equation}

\begin{lemma}\label{lem:dist-compare-ge2}
   For all $x,y \in \Omega$, it holds that
   \begin{align}
      d(x,y) &\leq p \cN^{\infty}(x-y) \sqrt{2 \tilde{\Lambda}_{\Omega}}\,.
   \end{align}
\end{lemma}

\begin{proof}
   Note that for $p \geq 2$ and any $a,b \geq 0$, it holds that
   \[
      |a^p - b^p| \leq p |a-b| \sqrt{a^{2(p-1)} + b^{2(p-1)}}\,.
   \]
   Therefore
   \[
      d(x,y) \leq p \max_{j \in [M]} \left(\left|\cN_j(x)-\cN_j(y)\right|\right) \left(\sum_{j=1}^M \cN_j(x)^{2(p-1)} + \cN_j(y)^{2(p-1)}\right)^{1/2}.\qedhere
   \]
\end{proof}

As in the previous section, we can use this in conjunction with \pref{cor:pre-diameter-bound} to bound the diameter, as
$\diam(\Omega, d) \lesssim p\sqrt{\tilde{\Lambda}_{\Omega}}\ \diam(B_{\bnorm}, \cN^\infty)$.

\begin{corollary}\label{cor:diameter-bound-ge2}
   For $p \geq 2$, it holds that $\diam(\Omega, d) \lesssim p \smoothconst \mathsf{w}_{\infty} \displaystyle \sqrt{\tilde{\Lambda}_{\Omega}}\,$.
\end{corollary}

What follows is the analogous entropy bound.

\begin{lemma}[Entropy bound]\label{lem:lp-chaining-geq2}
   It holds that
   \begin{align*}
      \gamma_2(\Omega, d) &\lesssim p \left(\smoothconst \mathsf{w}_{\infty} \psi_n \log(M) \log(n)\right) \sqrt{\tilde{\Lambda}_{\Omega}}
   \end{align*}
\end{lemma}

\begin{proof}
   Noting that both sides scale linearly in $\{\cN_j^p\}$, we may assume that $\tilde{\Lambda}_{\Omega} = 1$.
Applying \pref{lem:dist-compare-ge2} then gives
\begin{equation}\label{eq:d-compare-ge2}
   d(x,y) \lesssim p \cN^{\infty}(x-y)\,,\quad x,y \in \Omega\,.
\end{equation}

Applying \pref{lem:smooth-sudakov} with $N = \bnorm, \hat{N} = \cN^{\infty}$ yields
\[
   \sqrt{\log \cov(B_{\bnorm}, \cN^{\infty}, \e)} \lesssim \frac{\smoothconst}{\e} \int \cN^{\infty}(x)\,d\mu(x)
   \lesssim \frac{\smoothconst \psi_n \log M}{\e} \mathsf{w}_{\infty}\,,
\]
where the latter inequality is the first inequality in \pref{lem:tailbound}.

Thus defining $Q \seteq \psi_n \log(M) \mathsf{w}_{\infty}$,
we have
\begin{equation}\label{eq:pre-dud-1b}
   e_h(B_{\bnorm}, \cN^{\infty}) \lesssim 2^{-h/2} \smoothconst Q\,,\quad h \geq 0\,.
\end{equation}
Note that \pref{cor:diameter-bound-ge2} gives
\[
   \diam(B_{\bnorm}, \cN^{\infty}) \lesssim \smoothconst \mathsf{w}_{\infty}\,,
\]
and therefore using \pref{lem:norm-cover},
\begin{equation}\label{eq:pre-dud-2b}
   e_h(B_{\bnorm}, \cN^{\infty}) \lesssim \smoothconst \mathsf{w}_{\infty} e_h(B_{\cN^{\infty}}, \cN^{\infty}) \lesssim 2^{-2^h/n} \smoothconst \mathsf{w}_{\infty}\,,\quad h \geq 0\,.
\end{equation}
Using \eqref{eq:pre-dud-1b} and \eqref{eq:pre-dud-2b} in conjunction with the Dudley entropy bound \eqref{eq:dudley-v1} gives
\[
   \gamma_2(B_{\bnorm}, \cN^{\infty}) \lesssim \sum_{0 \leq h \leq 4 \log n} 2^{h/2} e_h(B_{\bnorm}, \cN^{\infty}) + \smoothconst \mathsf{w}_{\infty} \sum_{h > 4 \log n} 2^{h/2} 2^{-2^h/n} 
                    \lesssim \smoothconst Q \log n\,.
\]
The proof is complete since $\gamma_2(\Omega, d) \lesssim p\cdot \gamma_2(B_{\bnorm}, \cN^{\infty})$ by \eqref{eq:d-compare-ge2}.
\end{proof}

\subsection{Sparsification}

\newcommand{\genericnorm}{N}

We now complement the preceding entropy estimates with control of our desired sampling probabilities, beginning with a simple fact.

\begin{lemma}\label{lem:int-Nz}
   For any norm $\genericnorm$ on $\R^n$ and $p \geq 1$, if $\mu$ is the probability measure
   with $d\mu(x) \propto e^{-\genericnorm(x)^p}$, then
   \[
      \int \genericnorm(x)^p \,d\mu(x) = \frac{n}{p}
   \]
\end{lemma}

\begin{proof}
   Define $V(r) \seteq \vol_n\! \left(r B_{\genericnorm}\right) = r^n \vol_n(B_{\genericnorm})$, so $\frac{d}{dr}V(r) = n r^{n-1} \vol_n(B_{\genericnorm})$. Therefore,
\[
\int_{\R^n} \genericnorm(x)^p\,d\mu(x) =
 \frac{\int_{\R^n} \genericnorm(x)^p e^{-\genericnorm(x)^p}\,dx}{\int_{\R^n} e^{-\genericnorm(x)^p}\,dx} =
 \frac{\int_{0}^{\infty} r^p e^{-r^p} d V(r)}{\int_{0}^{\infty} e^{-r^p} d V(r)} =
 \frac{\int_{0}^{\infty} e^{-r^p} r^{n-1+p}\,dr}{\int_{0}^{\infty} e^{-r^p} r^{n-1}\,dr}\,.
\]
Make the subsitution $u = r^p$, yielding
\[
\int_{\R^n} \genericnorm(x)^p\,d\mu(x)
= \frac{\int_{0}^{\infty} e^{-u} u^{n/p}\,du}{\int_{0}^{\infty} e^{-u} u^{n/p-1}\,du} = \frac{n}{p}\,,
\]
where the latter equality follows from integration by parts.
\end{proof}

If $N$ is a semi-norm, we may apply \pref{lem:int-Nz} to the restriction of $N$ to $\ker(N)^{\perp}$,
yielding the following.

\begin{corollary}\label{cor:int-Nz}
   For any semi-norm $\genericnorm$ on $\R^n$ and $p \geq 1$, if $\mu$ is the probability measure on $\ker(\genericnorm)^{\perp}$
   with $d\mu(x) \propto e^{-\genericnorm(x)^p} dx$, then
   \[
      \int \genericnorm(x)^p \,d\mu(x) = \frac{n-\dim(\ker(N))}{p} \leq \frac{n}{p}\,.
   \]
\end{corollary}

Let us now fix $p \in [1,2]$, and consider semi-norms $N_1,\ldots,N_m$ on $\R^n$.
Define the semi-norm $N(x)$ by $N(x)^p \seteq N_1(x)^p + \cdots + N_m(x)^p$.
Suppose that $\bnorm$ is another semi-norm on $\R^n$ that is $p$-uniformly smooth with constant $\smoothconst$,
and that
\begin{equation}\label{eq:Kequiv}
   \bnorm(x) \leq N(x) \leq K \bnorm(x)\,\qquad \forall x \in \R^n\,.
\end{equation}
Let $\mu$ denote the probability measure whose density satisfies $d\mu(x) \propto e^{-\bnorm(x)^p} dx$, and
suppose that $\tau_1,\ldots,\tau_m \geq 0$ are numbers satisfying
\begin{equation}\label{eq:tau-assume}
   \E [N_i(\bm{Z})^p] \leq \tau_i \leq 2 \E[N_i(\bm{Z})^p]\,,
\end{equation}
where $\bm{Z}$ has law $\mu$. Define the probability vector $\rho \in \R^m_+$ by $\rho_i \seteq \tau_i/\|\tau\|_1$ for $i=1,\ldots,m$.

The next theorem establishes \pref{thm:l2-sparsification}.

\begin{theorem}
\label{thm:approx}
Let $p \in [1, 2]$. There is an explicit function $C : \Z_{+} \to \R_+$ such that $C(n) \lesssim (K\smoothconst \psi_n)^p (\log n)^2$ for
all $n \in \Z_+$, and so that for any $0 < \e < 1$ and $M \geq C(n) n (\log (n/\eps))^p \e^{-2}$,
the following holds.
If $i_1,\ldots,i_M$ are indicies sampled independently from $\rho$, then with probability at least $1-n^{-\Omega((\log n)^3)}$,
\[
 \left|N(x)^p - \frac{1}{M}\sum_{j\in[M]} \frac{N_{i_j}(x)^p}{\rho_{i_j}} \right| \le \eps N(x)^p\ , \quad \forall x \in \R^n\,.
\]
\end{theorem}

\begin{proof}
   Note that, by \eqref{eq:tau-assume},
\begin{equation}
   \label{eq:normalized}
\|\tau\|_1 \le 2\sum_{j=1}^M \int N_j(x)^p \,d\mu(x) = 2 \int N(x)^p\,d\mu(x) \le 2 K^p  \int \bnorm(x)^p\,d\mu(x) = 2K^p n / p,
\end{equation}
where the last equality follows from \pref{cor:int-Nz}.

Given $M \geq 1$ and $\nu \in [m]^M$, define the semi-norms $\cN_1,\ldots,\cN_M$ by
\[
   \cN_j(x)^p \seteq \frac{N_{\nu_j}(x)^p}{M \rho_{\nu_j}}\,,\qquad j=1,\ldots,M\,,
\]
and denote $\f_i(x) \seteq N_i(x)^p$ for $i=1,\ldots,m$
so that $d_{\rho,\nu}(x,y) = d(x,y)$, where $d_{\rho,\nu}$ is defined in \eqref{eq:d_rho_nu}, and $d$ is from \pref{def:entropy-objects}.

We will apply \pref{lem:sparsification-meta} with $F \seteq N$, $\Omega \seteq B_N$, and $\{\rho_i\},\{\f_i\}$ defined as above. To do so, we require bounds on
$\gamma_2(B_{N}, d_{\rho, \nu})$ and $\diam(B_{N}, d_{\rho,\nu})$.
Let us take $\Omega \seteq B_N$, noting that $B_N \subseteq B_{\bnorm}$ from \eqref{eq:Kequiv}.
Then \pref{lem:lp-chaining} yields 
\begin{equation}\label{eq:c-chaining}
   \gamma_2(B_N, d_{\rho, \nu}) = \gamma_2(B_N, d) \lesssim \left(\smoothconst \mathsf{w}_{\infty} \psi_n \log M\right)^{p/2} \log(n) \sqrt{\Lambda_{\Omega}}\,,
\end{equation}
Note that
\begin{align*}
\E[\cN_j(\bm{Z})^p]
    &=  \frac{1}{M \rho_{\nu_j}} \E[N_{\nu_j}(\bm{Z})^p]
    \stackrel{\eqref{eq:tau-assume}}{\leq} \frac{\tau_{\nu_j}}{M \rho_{\nu_j}}  = \frac{\norm{\tau}_1}{M}
\end{align*}
for all $j=1,\ldots,M$, and
therefore by monotonicity of $p$th moments,
\[
   \mathsf{w}_{\infty} \leq \max_{j \in [M]} \left(\E[\cN_j(\bm{Z})^p]\right)^{1/p} \leq \left(\frac{\|\tau\|_1}{M}\right)^{1/p}.
\]
By definition, it holds that
\[
   \Lambda_{\Omega} = \max_{x \in B_N} \sum_{j=1}^M \cN_j(x)^p = \max_{x \in B_N} \tilde{F}_{\rho,\nu}(x)\,.
\]
Substituting these bounds into \eqref{eq:c-chaining} gives
\begin{align*}
\gamma_2(B_{N}, d_{\rho, \nu}) &\lesssim M^{-1/2} \left(K \smoothconst \psi_n \log M\right)^{p/2} \log(n) \norm{\tau}_1^{1/2} \sqrt{\max_{x \in B_{N}}  \tilde{F}_{\rho,\nu}(x)} \\
& \stackrel{\eqref{eq:normalized}}{\lesssim} \frac{\left(K \smoothconst \psi_n \log M\right)^{p/2} \sqrt{n} \log n}{M^{1/2}} \sqrt{\max_{x \in B_{N}}  \tilde{F}_{\rho,\nu}(x)} \\
&\le \delta \sqrt{\max_{x \in B_{N}} \tilde{F}_{\rho,\nu}(x)}
\end{align*}
for some choice of $M$ sufficiently large and satisfying
\[
   M \lesssim n \delta^{-2} (\log n)^2 (\log(n/\delta))^p (K \smoothconst \psi_n)^p\,.
\]

In addition, \pref{cor:diameter-bound} and \eqref{eq:normalized} gives
\[
   \diam(B_N, d_{\rho,\nu}) \lesssim (\smoothconst \mathsf{w}_{\infty})^{p/2}  \sqrt{\Lambda_{\Omega}} \leq \left(\smoothconst^{p}  \frac{\norm{\tau}_1}{M}\right)^{1/2} \left(\max_{x \in B_{N}}  \tilde{F}_{\rho,\nu}(x)\right)^{1/2} \le \left(K^p\smoothconst^{p}  \frac{n}{M}\right)^{1/2} \left(\max_{x \in B_{N}}  \tilde{F}_{\rho,\nu}(x)\right)^{1/2}. 
\]
Using our choice of $M$ gives
\[
   \diam(B_N, d_{\rho,\nu}) \leq \frac{C_0 \delta}{(\log n)^{3/2}}\left(\max_{x \in B_{N}}  \tilde{F}_{\rho,\nu}(x)\right)^{1/2}
\]
for some universal constant $C_0 > 0$.

From \pref{lem:sparsification-meta} \eqref{eq:max-tail}, we conclude that for a universal constant $A > 0$ and any $0 \leq t \leq \frac{(\log n)^{3/2}}{2 C_0 A \delta}$,
\begin{equation}
    \Pr \left(\max_{x \in B_{N}} \left|N(x)^p - \frac{1}{M}\sum_{j=1}^M \frac{N_{i_j}(x)^p}{\rho_{i_j}} \right| > A \left(\delta + \frac{C_0 t \delta}{(\log n)^{3/2}}\right)\right)
    \leq e^{-At^2/4}\,.
\end{equation}

For $t \seteq \frac{(\log n)^{3/2}}{2 C_0 A}$, this shows that with probability at least $1-e^{- \Omega((\log n)^{3})}$,
\[
   \left|N(x)^p - \frac{1}{M}\sum_{j=1}^M  \frac{N_{i_j}(x)^p}{\rho_{i_j}} \right| \leq 2 A \delta N(x)^p\,,\quad  \forall x \in \R^n\,.
\]
Choosing $\delta \seteq \e/(2A)$ now yields the desired result.
\end{proof}

\subsubsection{Sparsification for $p \geq 2$}
We start by defining the sampling weights, analogous to \pref{lem:int-Nz}.
\begin{lemma}\label{lem:int-ge2}
   For any norm $\genericnorm$ on $\R^n$ and $p \geq 1$, if $\mu$ is the probability measure
   with $d\mu(x) \propto e^{-\genericnorm(x)^2}$, then
\[
   \int N(x)^p\,d\mu(x) \leq \left(\frac{n+p}{2}\right)^{p/2}.
\]
\end{lemma}

\begin{proof}
 As in the proof of \pref{lem:int-Nz}, 
write
\begin{align*}
   \frac{\int N(x)^p e^{-N(x)^2}\,dx}{\int e^{-N(x)^2}\,dx} = \frac{\int_0^{\infty} r^p e^{-r^2} dB(r)}{\int_0^{\infty} e^{-r^2} dB(r)}
   &= \frac{\int_0^{\infty} r^{p+n-1} e^{-r^2}\,dr}{\int_0^{\infty} r^{n-1} e^{-r^2}\,dr}\,.
\end{align*}
Now make the substitution $u = r^2$, so the left-hand side is
\[
   \frac{\int_0^{\infty} u^{(p+n)/2-1} e^{-u}\,dr}{\int_0^{\infty} u^{n/2-1} e^{-u}\,dr} = \frac{\Gamma((p+n)/2-1)}{\Gamma((n/2)-1)}\,,
\]
where we recall the definition of the $\Gamma$ function: For real $t \geq 0$,
\[
   \Gamma(t) = \int_0^{\infty} e^{-u} u^t\,du\,.
\]
Finally, note that $\Gamma(t+1)=t \Gamma(t)$ and $\Gamma(t+s) \leq t^s \Gamma(t)$ for all $0 < s < 1$ and $t \geq 0$ \cite{Wendel48}, hence
for $k \seteq \lfloor p/2\rfloor$,
\begin{align*}
   \frac{\Gamma((p+n)/2-1)}{\Gamma((n/2)-1)} &= \left(\frac{n+p}{2}-2\right) \left(\frac{n+p}{2}-3\right) \cdots \left(\frac{n+p}{2}-(k+1)\right)
   \frac{\Gamma(\frac{n+p}{2}-k-1))}{\Gamma(\frac{n}{2}-1)} \\
                                             &\leq
\left(\frac{n+p}{2}-2\right) \left(\frac{n+p}{2}-3\right) \cdots \left(\frac{n+p}{2}-(k+1)\right)\left(\frac{n}{2}-1\right)^{p/2-k} \\
                                             &\leq \left(\frac{n+p}{2}\right)^{p/2}.\qedhere
\end{align*}
\end{proof}

\begin{lemma}\label{cor:int-ge2}
   For any semi-norm $\genericnorm$ on $\R^n$ and $p \geq 1$, if $\mu$ is the probability measure
   on $\ker(\genericnorm)^{\perp}$
   with $d\mu(x) \propto e^{-\genericnorm(x)^2}\,dx$, then
\[
   \int N(x)^p\,d\mu(x) \leq \left(\frac{n+p}{2}\right)^{p/2}.
\]
\end{lemma}

\begin{theorem}
\label{thm:approx-ge2}
Suppose $p > 2$ and $N_1,\ldots,N_m$ are semi-norms on $\R^n$ such that the semi-norm defined by
$N(x)^p = N_1(x)^p + \cdots + N_m(x)^p$ is $K$-equivalent to a semi-norm $\bnorm$ that
is $2$-uniformly smooth with constant $\smoothconst$.
There is a weight vector $w \in \R_+^m$ with
\[
   |\supp(w)| \lesssim \frac{K^p \smoothconst^p}{\e^2} \left(\frac{n+p}{2}\right)^{p/2} (\psi_n \log(n/\e) \log(n))^2
\]
and such that
\[
   \left|N(x)^p - \sum_{i=1}^m w_i N_i(x)^p\right| \le \eps N(x)^p\ , \quad \forall x \in \R^n\,.
\]
\end{theorem}

\begin{proof}
   First, let us scale so that
   \[
      \bnorm(x) \leq N(x) \leq K \bnorm(x)\,,\qquad \forall x \in \R^n\,.
   \]
   Let $\mu$ denote the probability measure with $d\mu(x) \propto e^{-\bnorm(x)^2}$, and
define, for $i=1,\ldots,m$,
\begin{align*}
   \tau_i &\seteq \int N_i(x)^p\, d\mu(x) \\
   \rho_i &\seteq \frac{\tau_i}{\|\tau\|_1}\,.
\end{align*}
Note that \pref{cor:int-ge2} gives
\begin{equation}\label{eq:tau-bound}
   \|\tau\|_1 \leq K^p \left(\frac{n+p}{2}\right)^{p/2}.
\end{equation}

Given $M \geq 1$ and $\nu \in [m]^M$, define the semi-norms $\cN_1,\ldots,\cN_M$ by
\[
   \cN_j(x)^p \seteq \frac{N_{\nu_j}(x)^p}{M \rho_{\nu_j}}\,,
\]
and denote $\f_i(x) \seteq N_i(x)^p$ for $i=1,\ldots,m$
so that $d_{\rho,\nu}(x,y) = d(x,y)$, where $d_{\rho,\nu}$ is defined in \eqref{eq:d_rho_nu}, and $d$ is from \pref{def:entropy-objects}.

We will apply \pref{lem:sparsification-meta} with $F \seteq N$, $\Omega \seteq B_N$, and $\{\rho_i\},\{\f_i\}$ defined as above. To do so, we require a bound on
$\gamma_2(B_{N}, d_{\rho, \nu})$.
\pref{lem:lp-chaining-geq2} yields 
\begin{equation}\label{eq:c-chaining-ge2}
   \gamma_2(B_N, d_{\rho, \nu}) = \gamma_2(B_N, d) \lesssim p \left(\smoothconst \mathsf{w}_{\infty} \psi_n \log(M) \log(n)\right) \sqrt{\tilde{\Lambda}_{\Omega}}\,.
\end{equation}

Observe that
\begin{equation}\label{eq:wmax-bnd}
   \mathsf{w}_{\infty} = \max_{j \in [M]} \int \cN_j(x)\,d\mu(x) \leq \max_{j \in [M]} \left(\int \cN_j(x)^p \,d\mu(x)\right)^{1/p} \leq \left(\frac{\|\tau\|_1}{M}\right)^{1/p}.
\end{equation}
Thus from \pref{lem:lp-diam} and monotonicity of $p$th moments, we see that for $j=1,\ldots,M$,
\[
   \max_{x \in B_{\bnorm}} \cN_j(x)^p \lesssim \smoothconst^p \frac{\|\tau\|_1}{M}\,.
\]
So for $x \in B_{\bnorm}$, we can write
\[
   \sum_{j=1}^M \cN_j(x)^{2(p-1)} \lesssim \left(\smoothconst^p \frac{\|\tau\|_1}{M}\right)^{(p-2)/p} \sum_{j=1}^M \cN_j(x)^{p}\,.
\]
Recalling the definitions of $\Lambda_{\Omega}$ (\pref{def:entropy-objects}) and $\tilde{\Lambda}_{\Omega}$ \eqref{eq:tlam}, this gives
\[
   \sqrt{\tilde{\Lambda}_{\Omega}} \leq \left(\smoothconst^p \frac{\|\tau\|_1}{M}\right)^{1/2-1/p} \sqrt{\Lambda_{\Omega}}\,.
\]
Note also that
\[
   \Lambda_{\Omega} = \max_{x \in B_N} \sum_{j=1}^M \cN_j(x)^p = \max_{x \in B_N} \tilde{F}_{\rho,\nu}(x)\,.
\]
Thus in conjunction with \eqref{eq:tau-bound}, \eqref{eq:c-chaining-ge2}, and \eqref{eq:wmax-bnd}, we have
\[
   \gamma_2(B_N, d_{\rho,\nu}) \lesssim p \left(K^p \smoothconst^p M^{-1} \left(\frac{n+p}{2}\right)^{p/2}\right)^{1/2} \left( \psi_n \log(M) \log(n)\right) \left(\max_{x \in B_N} \tilde{F}_{\rho,\nu}(x)\right)^{1/2}\,.
\]

So for every $\e \in (0,1)$, there is a choice of
\[
   M \lesssim \frac{K^p \smoothconst^p p^2}{\e^2} \left(\frac{n+p}{2}\right)^{p/2} (\psi_n \log(n/\e) \log(n))^2
\]
such that
\[
   \gamma_2(B_N, d_{\rho,\nu}) \lesssim \e\left(\max_{x \in B_N} \tilde{F}_{\rho,\nu}(x)\right)^{1/2}.
\]
An application of \pref{lem:sparsification-meta} gives
\[
   \E_{\bm{\nu}} \max_{x \in B_N} \left|N(x)^p - \frac{1}{M} \sum_{j=1}^M \frac{N_{\bm{\nu}_j}(x)^p}{\rho_{\bm{\nu}_j}}\right| \lesssim \e\,.\qedhere
\]

\end{proof}

\subsection{Algorithms}
\label{sec:algorithm}

We first present an efficient algorithm for sampling in the case $p=1$.
Consider semi-norms $N_1,\ldots,N_m$ on $\R^n$ and suppose that each $N_i$ can be evaluated in time $\tEval$,
and that $N(x) \seteq N_1(x) + \cdots + N_m(x)$ is $(r,R)$-rounded for $0 < r \leq R$.

\begin{theorem}[Efficient sparsification]
   \label{thm:fast_sparse}
   If $N$ is $(r, R)$-rounded, then 
   for any $\e \geq n^{-O(1)}$, 
   there is an algorithm running in time $(m (\log n)^{O(1)} + n^{O(1)})(\log (mR/r))^{O(1)}\tEval$ that with high probability produces an
   $O(n \e^{-2} \log(n/\e) (\log n)^{2.5})$-sparse $\e$-approximation to $N$.
\end{theorem}

Suppose now that $\tilde{N}$ is a semi-norm on $\R^n$ that is $K$-equivalent to $N$, and let $\mu$
be the probability measure with density proportional to $e^{-\tilde{N}(x)}\,dx$.

\begin{lemma}[Sampling to sparsification]
\label{lem:samp_to_sparse}
For $h \geq 1$,
there is an algorithm that, given $O(h \psi_n \log(m+n))$ independent samples from $\mu$ and $\e > 0$,
computes with probability at least $1-(m+n)^{-h}$, an $s$-sparse $\e$-approximation to $N$ in time $O(m \psi_n \log(n+m) + s) \tEval$,
where $s \leq O(K^2 \e^{-2} n \log(n/\e) (\log n)^{2.5})$.
\end{lemma}

\begin{proof}
   Let $\bm{X}_1,\dots,\bm{X}_k \in \R^n$ be independent samples from $\mu$.
Denote, for $i=1,\ldots,m$,
\begin{align*}
   \tau_i &\seteq \frac{3}{2} \frac{1}{k} \left(N_i(\bm{X}_1)+N_i(\bm{X}_2)+\cdots+N_i(\bm{X}_k)\right) \\
   \sigma_i &\seteq \E[N_i(\bm{X}_1)]\,.
\end{align*}
Since $\mu$ is log-concave, \pref{cor:norm-concentration} asserts there is a constant $c > 0$ such that 
\[
   \Pr\left( \left|N_i(x_j) - \sigma_i \right|> t \right) \leq 2 \exp\left(- \frac{c t}{\psi_n \sigma_i}\right)
\]
Consequently, for some $k \lesssim h \psi_n \log(m+n)$, it holds that
\[
   \Pr\left(\sigma_i \leq \tau_i \leq 2 \sigma_i, i=1,\ldots,m\right) \geq 1 - (m+n)^{-h}\,.
\]
Thus with high probability, \eqref{eq:tau-assume} it satisfied for $p=1$,
and one obtains the desired sparse approximation using \pref{thm:approx} with $p=1$.
\end{proof}

The preceding lemma shows that sampling from a distribution with $d\mu(x) \propto e^{-\tilde{N}(x)}$ suffices to efficiently sparsify a semi-norm
$N$ that is $K$-equivalent to $\tilde{N}$.
A long line of work establishes algorithms that sample from a distribution
that is close to uniform on any well-conditioned convex body $A \subseteq \R^n$, given
only membership oracles to $A$.
In the following statement, let $B_2^n$ denote the Euclidean unit ball in $\R^n$.

\begin{theorem}[{\cite[Theorem 1.5]{JLLV21}}, {\cite[Theorem 1.2]{CV18}}]
\label{thm:sample}
There is an algorithm that, given a convex body $A \subseteq \R^n$ satisfying $r \cdot B_2^n \subseteq A \subseteq R \cdot B_2^n$
and $\e > 0$,
samples from a distribution that is within total variation distance $\eps$ from the uniform measure on $A$ 
using $O(n^3 (\log \frac{nR}{\e r})^{O(1)})$ membership oracle queries to
$A$, and $(n (\log \frac{nR}{\e r}))^{O(1)}$ additional time.
\end{theorem}

When $N$ is a norm,
one obtains immediately an algorithm for sampling from the measure $\mu$ on $\R^n$ with density 
$d\mu(x) \propto e^{-N(x)}\,dx$ using evaluations of $N(x)$.

\begin{corollary}
\label{cor:sample}
There is an algorithm that, given an $(r,R)$-rounded norm $N$ on $\R^n$
and $\e > 0$, samples
from a distribution that is within total variation distance $\e$ from the measure $\mu$ with density proportional to $e^{-N(x)}\,dx$
using $O(n^3 (\log \frac{nR}{\e r})^{O(1)})$ evaluations of $N(x)$, and
$(n (\log \frac{nR}{\e r}))^{O(1)}$ additional time.
\end{corollary}

\begin{proof}
Note that if $\bm{Z}$ has law $\mu$, then the density of $N(\bm{Z})$ is proportional to $e^{-\lambda} \lambda^{n-1}$.
In other words, $N(\bm{Z})$ has the law of a sum of $n$ i.i.d. exponential random variables.
Let $\bm{\lambda}$ be a sample from the latter distribution.
The algorithm is as follows: Sample a point $\bm{X}$ from the uniform measure on $B_N$ using \pref{thm:sample},
and then output the point $\bm{\lambda} \bm{X}/N(\bm{X})$.
\end{proof}

Combining \pref{lem:samp_to_sparse} and \pref{cor:sample}, we see that if
one can sample from the distribution induced by a sparsifier, then one can efficiently
sparsify and if one can efficiently sparsify, then one can can perform the requisite
sampling.

This chicken-and-egg problem has arisen for a variety of
sparsification problems and there is a relatively simple and standard solution
introduced in \cite{mp12rowSampArxivV1} that has been used in a range of
settings; see e.g., \cite{KapralovLMMS17, jss18arxiv, AhmadinejadJSS19}).

Instead of simply sampling proportional to $e^{-N(x)}$ directly, we first sample proportional
to the density $\exp(- (N(x) + t \norm{x}_2))$, where $t$ is chosen large enough that the sampling problem is trivial.
This gives a sparsifier for  $N(x) + t \norm{x}_2$ which, in turn, can be used to efficiently sparsify
$N(x) + t/2 \norm{x}_2$. Iterating allows us to establish \pref{thm:fast_sparse}.

\begin{proof}[Proof of \pref{thm:fast_sparse}]
   Recall our assumption that $r \|x\|_2 \leq N(x) \leq R \|x\|_2$ for all $x \in \ker(N)^{\perp}$.
For $t \geq 0$, denote $N_t(x) \seteq N(x) + t \|x\|_2$.
Note that $N_R$ is $2$-equivalent to $R \|x\|_2$, and consequently by sampling from $d\mu(x) \propto \exp(-R\|x\|_2)$ using \pref{cor:sample}, we can use \pref{lem:samp_to_sparse}
to obtain an $\tilde{O}(n)$-sparse $1/2$-approximation to $N_R$.

Now for any $t \in [\e r, R]$, suppose $\tilde{N}_t$ is an $\tilde{O}(n)$-sparse $1/2$-approximation to $N_t$.
Note that $\tilde{N}_t$ is $(t/2, 4R)$-rounded. Thus, using \pref{cor:sample}, we can compute a sample 
from the distribution with density $\propto e^{-\tilde{N}_t(x)}$ in time $(n \log(R/r))^{O(1)} \tEval$. We can ignore the total variation error in \pref{cor:sample} as long as it is less than $m^{-O(1)}$ and charge it to the failure probability.
Since $N_{t/2}$ is $2$-equivalent to $N_t$, which is $2$-equivalent to $\tilde{N}_t$,
we can use \pref{lem:samp_to_sparse} to obtain an $\tilde{O}(n)$-sparse $1/2$-approximation to $N_{t/2}$.

After $O(\log (R/(\e r)))$ iterations, one obtains an $\tilde{O}(n)$-sparse $1/2$-approximation to $N_{\e r}$. A final application of \pref{lem:samp_to_sparse} obtains an $O(n \e^{-2} \log(n/\e) (\log n)^{2.5})$-sparse
$\e$-approximation to $N_{\e r}$. To conclude, note that for all $x \in \ker(N)^{\perp}$, $N_{\e r}$ is $(1+\e)$-equivalent to $N$.
Moreover, in $N_{\e r}(x) = N(x) + \e r \|x\|_2$, only the summand $\e r \|x\|_2$ fails to vanish on $\ker(N)$. This can be removed from $N_{\e r}$ to obtain
a $(1+2\e)$-approximation to $N$ with the same sparsity. The result then follows by applying this procedure with a smaller value of $\epsilon$.
\end{proof}

\begin{remark}[Algorithm for $1 < p \leq 2$]
We note that it is possible to extend \pref{thm:fast_sparse} to the setting of $1 < p \leq 2$ under a mild additional assumption.
Specifically, we need to assume that each semi-norm $N_i$ is itself $K$-equivalent
to a $p$-uniformly smooth semi-norm $\bnorm_i$ with constant $\cS_p$, and that we have oracle access to $\bnorm_i$.

For any weights $w_1,\ldots,w_m \geq 0$,
the semi-norm $N_w(x) \seteq \left(w_1 N_1(x)^p + \cdots + w_m N_m(x)^p\right)^{1/p}$
is then $K$-equivalent to the semi-norm $\bnorm_w(x) \seteq \left(w_1 \bnorm_1(x)^p + \cdots + w_m \bnorm_m(x)^p\right)^{1/p}$,
where each $\bnorm_i$ is $p$-uniformly smooth with constant $\cS_p$.
 Since the $\ell_p$ sum of $p$-uniformly smooth semi-norms is also $p$-uniformly smooth quantitatively (see \cite{Figiel76}),
it holds that $N_w$ is $K$-equivalent to a semi-norm $\bnorm_w$ that is $p$-uniformly smooth with constant $O(\cS_p)$.
One can then proceed along similar lines using the interpolants
\[
   N_t(x) \seteq \left(N(x)^p + t \|x\|_2^p\right)^{1/p}\,,
\]
which are similarly $K$-equivalent to the $p$-uniformly smooth norm $\bnorm_t(x) = \left(\bnorm(x)^p + t \|x\|_2^p\right)^{1/p}$,
since $\|\cdot\|_2$ is $p$-uniformly smooth with constant $1$ for
any $1 \leq p \leq 2$.
\end{remark}

\subsubsection{Sparsifying symmetric submodular functions}
\label{sec:sfm}

First recall that the Lov\'asz extension $\bar{F}$ is a semi-norm. This follows because $\bar{F}$ can be expressed as
\[ \bar{F}(x) = \int_{-\infty}^{\infty} F(\{i : x_i \le t\})\,dt. \]
Note that the integral is finite because $F(\emptyset) = F(V) = 0$, and
clearly $\bar{F}(cx) = c\bar{F}(x)$ for all $c > 0$.
Also because $F$ is symmetric we have $F(x)=\int_{-\infty}^{\infty} F(\{i : x_i \leq t\})\,dt = \int_{-\infty}^{\infty} F(\{i : x_i \geq t\})\,dt = F(-x)$.
Finally, it is a standard fact that $F$ is submodular if and only if $\bar{F}$ is convex. Thus, $\bar{F}$ is indeed a semi-norm.

\begin{proof}[Proof of \pref{cor:sfm}]
We assume that $\eps \ge m^{-1/2}$, else the desired sparsity bound is trivial.

Let $\bar{f}_1,\ldots,\bar{f}_m$ denote the respective Lov\'asz extensions of $f_1,\ldots,f_m$, and let $\bar{F}$ denote the
Lov\'asz extension of $F$.
Define
$\tilde{F}(x) \seteq \bar{F}(x) + m^{-4}\|x\|_2$ and $\tilde{f}_i(x) \seteq \bar{f}_i(x) + m^{-5}\|x\|_2$ so
that $\tilde{F}(x) = \tilde{f}_1(x)+\cdots+\tilde{f}_m(x)$. Clearly each $\tilde{f}_i$ is $(m^{-5}, O(nR))$-rounded as $\tilde{f}_i(x) \le 2\|x\|_\infty R \le 2R\sqrt{n} \|x\|_2$. Thus \pref{thm:fast_sparse} yields weights $w \in \R_+^m$ with the asserted sparsity bound and such that
\[ \left| \tilde{F}(x) - \sum_{i=1}^m w_i \tilde{f}_i(x) \right| \le \eps \tilde{F}(x)\,,\quad \forall  x \in \R^n\,. \]

Additionally, the unbiased sampling scheme of \pref{sec:symmetrization} guarantees that $\E[w_1+\cdots+w_m] = m$, so $\sum_{i=1}^m w_i \le 2m$ with probability
at least $1/2$.
Assuming this holds, let us argue that $\left| F(S) - \sum_{i \in [m]} w_i f_i(S) \right| \le 2\eps F(S)$ for all $S \subseteq V$. Indeed,
\begin{align*} \left| F(S) - \sum_{i=1}^m w_i f_i(S) \right| \le \eps\tilde{F}(S) + \left(m + \sum_{i=1}^m w_i\right) m^{-5}\|x\|_2 \le \eps F(S) + m^{-3}\,.
\end{align*}
This is at most $2\eps F(S)$ if $F(S) \ge 1$, since we assumed that $\e \geq m^{-1/2}$.

If, on the other hand, $F(S) = 0$, then we conclude that all $f_i(S) = 0$ for all $i \in \supp(w)$.
This is because the weights given by the independent sampling procedure (recall \pref{sec:symmetrization})
are at least $1/M \geq 1/m$, and each function $f_i$ is integer-valued. Thus $w_1 f_1(S) + \cdots + w_m f_m(S) = 0$ as well.
\end{proof}

\section{Lewis weights}
\label{sec:sumsmooth}

When working with a subspace of $L_p(\mu)$, it is often useful to 
perform a ``change of density'' in order to compare the $L_p$ norm to some other norm
(in the present setting, to an $L_2$ norm); see, e.g., \cite{JS01}.
A classical paper of Lewis \cite{Lewis78} describes a very useful change
of density that has applications to sparsification problems like $\ell_p$ row sampling \cite{CP15}
and dimension reduction for linear subspaces of $\ell_p$ \cite{BLM89, Talagrand95}.

Let $\alpha$ be a norm on the space of linear operators $\R^n \to \R^n$.
Following Lewis \cite{Lewis79},
one can consider the corresponding optimization
\begin{equation}\label{eq:psi-opt}
   \max \left\{\,|\!\det(U)| : \alpha(U) \leq 1 \right\}.
\end{equation}

As one example, suppose that $K \subseteq \R^n$ is a symmetric convex body
and $\|x\|_K$ is the associated norm.
Define the operator norm
\[
   \alpha(U) \seteq \max_{\|x\|_2 = 1} \|U x\|_K.
\]
For an optimizer $U^{\star}$ of \eqref{eq:psi-opt}, $U^{\star}(B_2^n)$
is the John ellipsoid of $K$, i.e.,
a maximum volume ellipsoid such that $U^{\star}(B_2^n) \subseteq K$.
Indeed: $\alpha(U) \leq 1 \iff U(B_2^n) \subseteq K$, and the volume
of $U(B_2^n)$ is proportional to $\det(U)$.

\paragraph{$\ell_p$ Lewis weights for a matrix $A$}
Consider now a linear operator $A : \R^n \to \R^k$, and let
$a_1,\ldots,a_m$ be the rows of $A$.
Define for any $1 \leq p < \infty$, the norm
\[
   \alpha(U) \seteq \left(\sum_{i=1}^k \|U a_i\|_2^{p}\right)^{1/p},
\]
We remark that this $\alpha$ coincides with the absolutely $p$-summing operator norm when
$U$ is considered as an operator $U : E \to F$ with $F \seteq A(\R^n) \subseteq \ell_p^k$,
and where $E$ is $\R^n$ equipped with the Euclidean norm $x \mapsto \|(A^{\top} A)^{-1/2} x\|_2$.
See, e.g., \cite{DHT95} for background on absolutely summing operators.

For $1 \leq p \leq 2$,
the optimality condition for \eqref{eq:psi-opt} yields the existence of
a nonnegative diagonal matrix $W$ such that $\alpha((A^{\top} W A)^{-1/2}) \leq n^{1/p}$, and
\[
   \max_{i \in [k]} \frac{|\langle a_i,x\rangle|}{\|(A^{\top} W A)^{-1/2} a_i\|_2} \leq \|(A^{\top} W A)^{1/2} x\|_2 \leq \|Ax\|_p\,.
\]
In particular, this bounds the contribution of every coordinate to the $\ell_p$ norm: Denoting
$\alpha_i \seteq \|(A^{\top} W A)^{-1/2} a_i\|_2$, we have
\[
   |\langle a_i, x\rangle| \leq \alpha_i \|Ax\|_p\,,\quad \forall x \in \R^n\,,
\]
and $\alpha^p_1+\cdots+\alpha^p_k = \alpha((A^{\top} W A)^{-1/2})^p = n$.
The diaognal entries of $W$ are often referred to as $\ell_p$ Lewis weights \cite{CP15}.

\paragraph{Block $\ell_{\infty}$ weights}
In order to construct spectral sparsifiers for hypergraphs, the authors
of \cite{KKTY21b} implicitly consider the following setting, couched
in the language of effective resistances on graphs.
Suppose $S_1 \cup \cdots \cup S_m = [k]$ forms a partition of the index set,
and define
\begin{align*}
   \alpha(U) &\seteq \left(\sum_{j=1}^m \max_{i \in S_j} \|U a_i\|_2^2\right)^{1/2}.
\end{align*}
Their construction was clarified and extended in \cite{Lee23,JLS23}.
We now present a substantial generalization that will be a useful tool in proving \pref{thm:sos-lp} and \pref{thm:sos-general}.

\begin{definition}[Block norm]
   \label{def:block-norm}
Consider any $p_1,\ldots,p_m,q \in [1,\infty]$, and a partition $S_1 \cup \cdots \cup S_m = [k]$.
For $p_j < \infty$, define 
\begin{align*}
   \cN_j(u) &\seteq \left(\sum_{i \in S_j} |u_i|^{p_j} \right)^{1/p_j}\,,
\end{align*}
and for $p_j = \infty$, take $\cN_j(u) \seteq \max \{ |u_i| : i \in S_j \}$.
Define $\cN(u) \seteq \left\|(\cN_1(u),\ldots,\cN_m(u))\right\|_q$.
\end{definition}

Throughout, we let $\cL(\R^n,\R^k)$ denotes the linear space of linear operators from $\R^n$ to $\R^k$, and denote $\cL(\R^n) = \cL(\R^n, \R^n)$. We prove the next lemma in \pref{sec:block-lewis}.

\begin{lemma}\label{lem:lp-lewis}
   Consider $p_1,\ldots,p_m \in [2,\infty]$ and $q \in [1,\infty)$.
   Let $\cN_1,\ldots,\cN_m$ and $\cN$ be as in \pref{def:block-norm}.
   Fix $A \in \cL(\R^n,\R^k)$ with $\rank(A)=n$, and denote for $j=1,\ldots,m$,
   \[
      \alpha_j(U) \seteq \cN_j\!\left(\|U A^{\top} e_1\|_2,\ldots,\|U A^{\top} e_k\|_2\right)
   \]
   Then there is a nonnegative
   diagonal matrix $W$ such that for $U = (A^{\top} W A)^{-1/2}$, the following are true:
\begin{enumerate}[label=\textrm{(\arabic*)}]
   \item \label{enumlew:a} 
      It holds that
      \[
         \alpha_1(U)^q + \cdots + \alpha_m(U)^q = 
         \begin{cases}
            n & 1 \leq q \leq 2 \\
            n^{q/2} & q \geq 2\,.
         \end{cases}
      \]
   \item \label{enumlew:b} For all $x \in \R^n$,
         \[
            \cN_j(A x) \leq \alpha_j(U) \|U^{-1} x\|_2 \leq \alpha_j(U) \cN(A x)\,.
         \]
   \end{enumerate}
\end{lemma}

\subsection{Block Lewis weights}
\label{sec:block-lewis}

For a norm $\cN$ on $\R^k$ and $A \in \cL(\R^n,\R^k)$, define the norm $\alpha_{\cN,A}$ on $\cL(\R^n)$ by
\begin{equation}\label{eq:normdef}
   \alpha_{\cN,A}(U) \seteq \cN\!\left(\|U A^{\top} e_1\|_2,\ldots,\|U A^{\top} e_k\|_2\right)\,.
\end{equation}

We consider the optimization \eqref{eq:psi-opt}.
As observed in \cite{SZ01}, the analysis of \eqref{eq:psi-opt} does not rely on duality in a fundamental way.

\begin{lemma}
   \label{lem:opt-analysis}
   If $\cN$ is continuously differentiable,
   then there is an invertible, self-adjoint $U \in \cL(\R^n)$ such that 
   $\alpha_{\cN,A}(U) = 1$, and
   \[
      U = (A^{\top} \gamma W A)^{-1/2}\,,
   \]
   where $W$ is the diagonal matrix with
   \begin{equation}\label{eq:diag-scaling}
      W_{ii} = \frac{\partial_{x_i} \cN\!\left(\|U A^{\top} e_1\|_2,\ldots,\|U A^{\top} e_k\|_2\right)}{\|U A^{\top} e_i\|_2}\,,\quad i=1,\ldots,k\,,
   \end{equation}
   and
   \begin{equation}\label{eq:gamma-val}
      \gamma = \left(\frac{1}{n} \sum_{i=1}^k \|U A^{\top} e_i\|_2\ \partial_{x_i} \cN\!\left(\|U A^{\top} e_1\|_2,\ldots,\|U A^{\top} e_k\|_2\right)\right)^{-1}.
   \end{equation}
\end{lemma}

\begin{proof}
   Define the operator $B \in \cL(\R^n,\R^k)$ by
   \[
      B_{uv} \seteq \partial_{U_{uv}} \alpha(U) =
      \sum_{i=1}^k \frac{\partial_{x_i} \cN\left(\|U A^{\top} e_1\|_2,\ldots,\|U A^{\top} e_k\|_2\right)}{\|U A^{\top} e_i\|_2} (UA^{\top})_{ui} A_{iv}\,.
   \]
   Note that
   \[
      B = U A^{\top} W A\,,
   \]
   where $W$ is the diagonal matrix defined in \eqref{eq:diag-scaling}.
   
   Next, observe that for any invertible $U \in \cL(\R^n)$,
   \[
      \partial_{U_{uv}} \det(U) = \det(U) (U^{-1})_{vu}\,.
   \]
   Hence if $U_0$ is an optimal solution to \eqref{eq:psi-opt} with $\alpha = \alpha_{\cN,A}$, then
   \[
      U_0 A^{\top} W A = \lambda U_0^{-\top}
   \]
   where $\lambda > 0$ is the Lagrange multiplier corresponding to the constraint $\alpha(U_0) \leq 1$.

   Let us take $U \seteq (U_0^{\top} U_0)^{1/2}$ so that $\alpha(U)=\alpha(U_0)=1$.
   To compute the value of $\lambda$, use $A^{\top} W A = \lambda (U_0^{\top} U_0)^{-1} = \lambda U^{-2}$ to write
   \[
      \lambda n = \lambda \tr(U^2 U^{-2}) = \tr(U^2 A^T W A) = \sum_{i=1}^k W_{ii} \tr(U^2 A^{\top} e_i e_i^{\top} A) = \sum_{i=1}^k W_{ii} \|U A^{\top} e_i\|_2^2\,.
   \]
   Using again the definition \eqref{eq:diag-scaling}, we have
   \[
      \lambda = \frac{1}{n} \sum_{i=1}^k W_{ii} \|U A^{\top} e_i\|_2^2 = \frac{1}{\gamma}\,,
   \]
   showing that $U = (A^{\top} (\gamma W) A)^{-1/2}$.
\end{proof}

\begin{lemma}\label{lem:2-hold}
   Suppose that $\cN$ and $\cN_1,\ldots,\cN_m$ are as in \pref{def:block-norm} with $p_j \in [2,\infty]$ for $j=1,\ldots,m$,
   and $q \in [1,\infty)$.
   Then for any $A \in \cL(\R^n,\R^k)$, 
there is a diagonal matrix $W$ such that
$U = (A^{\top} W A)^{-1/2}$ satisfies $\alpha_{\cN,A}(U)^q \leq 1$, and for $j \in [m]$ and $i \in S_j$,
\begin{equation}\label{eq:wii}
   W_{ii} = 
\begin{cases}
   n u_i^{p_j-2} \cN_j(u)^{q-p_j} & p_j < \infty \\
   n v_i \cN_j(u)^{q-1} & p_j = \infty\,,
\end{cases}
\end{equation}
for some $v \in \R^k$ such that $\cN_j^*(v) = \cN_j(u)^{-1}$ when $p_j=\infty$.
\end{lemma}

\begin{proof}
   Let $U \in \cL(\R^n)$ be the map guaranteed by \pref{lem:opt-analysis} applied with the $\cN$ and $A$.
   (By a simple approximation argument, we may assume that $\cN$ is $\cC^1$ for the cases
   where $q=1$ or $p_j = \infty$.)
   
   Note that for $j \in [m]$ and $i \in S_j$, if $p_j < \infty$, then
   \[
      \partial_{u_i} \cN(u_1,\ldots,u_k) = \sign(u_i) |u_i|^{p_j-1} \cN_j(u)^{q-p_j} \cN(u)^{1-q}\,,
   \]
   and otherwise, for $p_j=\infty$,
   we must consider the collection of subgradients $v \in \R^{k}$ with $\cN_j^*(v) = \cN_j(u)^{-1}$.

   Defining $u_i \seteq \|U A^{\top} e_i\|_2$ for $i=1,\ldots,k$
   and plugging this into \eqref{eq:diag-scaling} gives,
   for $j \in [m]$ and $i \in S_j$,
\[
   W_{ii} = 
   \begin{cases}
       u_i^{p_j-2} \cN_j(u)^{q-p_j} & p_j < \infty \\
       v_i \cN_j(u)^{q-1} & p_j = \infty\,,
   \end{cases}
   \]
   where we have used the fact that $\cN(u) = \alpha_{\cN,A}(U) = 1$.

Now compute
\begin{align*}
\sum_{i=1}^k \|U A^{\top} e_i\|_2\ \partial_{x_i} \cN\!\left(\|U A^{\top} e_1\|_2,\ldots,\|U A^{\top} e_k\|_2\right)
= \sum_{j=1}^m \cN_j(u)^{q-p_j} \cN_j(u)^{p_j}
= 1\,.
\end{align*}
From \eqref{eq:gamma-val} we conclude that $\gamma = n$.
\end{proof}

Let us now use this to prove \pref{lem:lp-lewis}.

\begin{proof}[Proof of \pref{lem:lp-lewis}]
   Let $U = (A^{\top} W A)^{-1/2} \in \cL(\R^n)$ be the operator guaranteed by \pref{lem:2-hold} applied with norm $\cN$ and $A \in \cL(\R^n,\R^k)$.
   Denote $a_i \seteq A^{\top} e_i$ for $i=1,\ldots,k$ and the vector
$u\in \R^k$ by $u_i=\|U a_i\|_2$ for $i=1,\ldots,k$.
Recall that $\alpha_{\cN,A}(U)^q = \alpha_1(U)^q + \cdots + \alpha_m(U)^q = 1$.

Let us first perform the transformation $W \mapsto n^{-2/q} W$ and $U \mapsto n^{1/q} U$
so that $\alpha_1(U)^q + \cdots + \alpha_m(U)^q = n$ and, consulting \eqref{eq:wii},
define, for $j \in [m]$ and $i \in S_j$,
\[
   w_i \seteq W_{ii} = 
\begin{cases}
    u_i^{p_j-2} \cN_j(u)^{q-p_j} & p_j < \infty \\
    v_i  \cN_j(u)^{q-1} & p_j = \infty\,,
\end{cases}
\]
where $v \in \R^k$ is such that $\cN_j^*(v) = \cN_j(u)^{-1}$ when $p_j=\infty$.

Now use
$|\langle a_i,x\rangle| \leq \|U a_i\|_2 \|U^{-1} x\|_2$ to bound, for $p_j < \infty$,
\begin{align}
   \cN_j(Ax) = \left(\sum_{i \in S_j} |\langle a_i,x\rangle|^{p_j}\right)^{1/p_j} &\leq \|U^{-1} x\|_2 \left(\sum_{i \in S_j} \|U a_i\|^{p_j}_2 \right)^{1/p_j} \nonumber \\
                                                                                  &= \|U^{-1} x\|_2\, \alpha_j(U) = \|U^{-1} x\|_2\, \cN_j(u)\,,\label{eq:Nj}
\end{align}
verifying the first inequality in \pref{enumlew:b}. Clearly the same argument applies for $p_j=\infty$ as well.

Let $w_{S_j} \in \R^{S_j}$ denote the vector with $(w_{S_j})_i = w_i$.
We will abuse notation slightly and let $\|w_{S_j}\|_{p_j/(p_j-2)}$ denote $\|w_{S_j}\|_{\infty}$ when $p_j=2$ and denote $\|w_{S_j}\|_1$ when $p_j=\infty$.
For $2 < p_j < \infty$, we have
\begin{equation}\label{eq:wh}
   \|w_{S_j}\|_{p_j/(p_j-2)} = \left(\sum_{i \in S_j} u_i^{p_j}\right)^{(p_j-2)/p_j} \cN_j(u)^{q-p_j} = \cN_j(u)^{p_j-2} \cN_j(u)^{q-p_j}= \cN_j(u)^{q-2}\,.
\end{equation}
This remains true in the other cases as well:
When $p_j=2$, $\|w_{S_j}\|_{\infty} = \cN_j(u)^{q-2}$.
When $p_j=\infty$, $\|w_{S_j}\|_1 = \cN^*_j(v) \cN_j(u)^{q-1} = \cN_j(u)^{q-2}$.

Note that $\|U^{-\top} x\|_2^2 = \langle x, (U^{\top} U)^{-1} x\rangle$ and $(U^{\top} U)^{-1} = A^{\top} W A$.
Using this and H\"older's inequality with exponents $p_j/(p_j-2)$ and $p_j/2$, write
\begin{equation}\label{eq:e2a}
   \|U^{-\top} x\|_2^2 = \sum_{i=1}^k w_i |\langle a_i,x\rangle|^2 
                                                \leq 
                                                \sum_{j=1}^m \|w_{S_j}\|_{p_j/(p_j-2)}
                                                   \left(\sum_{i \in S_j} |\langle a_i,x\rangle|^{p_j}\right)^{2/p_j}  =
                                                \sum_{j=1}^m \cN_j(u)^{q-2} \cN_j(Ax)^2\,,
\end{equation}
where the last equality uses \eqref{eq:wh}.

\medskip
\noindent
{\bf Case I:} $1 \leq q \leq 2$.

\smallskip

To verify \pref{enumlew:a}, note that
\[
   \sum_{j=1}^m \alpha_j(U)^q = \cN(u)^q = n\,.
\]
Now observe that \eqref{eq:Nj} gives
\[
   \cN_j(u)^{q-2} \cN_j(Ax)^2 \leq \cN_j(Ax)^q \|U^{-1} x\|_2^{2-q}\,.
\]
In conjunction with \eqref{eq:e2a}, this yields
\[
   \|U^{-1} x\|_2^2 \leq \|U^{-1} x\|_2^{2-q} \sum_{j=1}^m \cN_j(Ax)^q\,,
\]
and simplifying yields
\[
   \|U^{-1} x\|_2 \leq \cN(Ax)\,,
\]
verifying the second inequality in \pref{enumlew:b}.

\medskip
\noindent
{\bf Case II:} $q > 2$.

\smallskip

Use H\"older's inequality with exponents $q/(q-2)$ and $q/2$ in \eqref{eq:e2a} to bound
\[
   \|U^{-1} x\|_2^2 \leq \left(\sum_{j=1}^m \cN_j(u)^q\right)^{(q-2)/q} \left(\sum_{j=1}^m \cN_j(Ax)^q\right)^{2/q} \leq n^{1-2/q} \cN(Ax)^2\,.
\]
Now replacing $U$ by $n^{1/2-1/q} U$ gives $\|U^{-1} x\|_2 \leq \cN(Ax)$ and $\alpha_1(U)^q + \cdots + \alpha_m(U)^q = n^{q/2}$.
\end{proof}

\subsubsection{Arbitrary norms}
\label{sec:general-infinity}

The $p_1 = \cdots = p_m =\infty$ case of \pref{lem:lp-lewis} gives a consequence
for {\em any} collection $N_1,\ldots,N_m$ of norms on $\R^n$ by embedding
them into a subspace of $\ell_{\infty}$.

\begin{lemma}\label{lem:block-lewis}
   For any $1 \leq q < \infty$,
there are numbers $\alpha_1,\ldots,\alpha_m \geq 0$ and a linear transformation $U : \R^n \to \R^n$ such that
\[
   \alpha_1^q + \cdots + \alpha_m^q \leq
   \begin{cases}
         n & 1 \leq q \leq 2 \\
         n^{q/2} & q \geq 2\,,
   \end{cases}
\]
	and for every $j \in [m]$ and $x \in \R^n$,
	\begin{align*}
      N_j(U x) &\leq \alpha_j \|x\|_2 \leq \alpha_j \left(\sum_{j=1}^m N_j(U x)^q\right)^{1/q}\,.
	\end{align*}
\end{lemma}

\begin{proof}
Let $N^*_j$ denote the dual norm to $N_j$.
Then for every $\delta > 0$ there is a finite set $V_{\delta} \subseteq B_{N_j^*}$ such that
\[
   N_j(x) = \sup_{N_j^*(y) \leq 1} \langle x,y\rangle \geq (1 - \delta) \max_{y \in V_{\delta}} \langle x,y\rangle = (1-\delta) \|A_j x\|_{\infty}
\]
for some map $A_j : \R^n \to \R^{|V_{\delta}|}$. 

Let $U_{\delta} : \R^n \to \R^n$ be the transformation
guaranteed by \pref{lem:lp-lewis} applied to the operator defined by
$A(x) = A_1(x) \oplus \cdots \oplus A_m(x)$.
This yields weights $\alpha_{1,\delta},\ldots,\alpha_{m,\delta} \geq 0$ such that
\[
   \sum_{j=1}^m \alpha_{j,\delta}^q = 
   \begin{cases}
         n & 1 \leq q \leq 2 \\
         n^{q/2} & q \geq 2\,,
   \end{cases}
\]
   and for all $x \in \R^n$,
   \[
      N_j(U_{\delta} x) \leq \alpha_{j,\delta} \|x\|_2 \leq (1-\delta)^{-1} \alpha_{j,\delta} N(U_{\delta} x)\,,
   \]
   where $N(x) \seteq \left(\sum_{j=1}^m N_j(x)^q \right)^{1/q}$.

It holds that $N(U_{\delta} x) \leq \sqrt{n} \|x\|_2$ for all $x \in \R^n$, and therefore
as long as $N$ is a genuine norm on $\R^n$ (and not simply a semi-norm), the family $\{U_{\delta} : \delta > 0\}$
is contained in a compact subset of $\cL(\R^n)$. In the case that $N$ is not a genuine norm,
one can simply restrict $U_{\delta}$ to $\ker(N)^{\perp}$ and argue there.

Define $s_{\delta} \seteq (U_{\delta}, \alpha_{1,\delta},\ldots,\alpha_{m,\delta})$,
and then the preceding bounds similarly show that $\{ s_{\delta} : \delta > 0 \}$ lies in a compact subset of $\cL(\R^n) \times \R^m$, and
therefore contains an accumulation point $(U,\alpha_1,\ldots,\alpha_m)$ satisfying the desired conclusion.
\end{proof}

\subsection{Sums of powers of general norms}

\newcommand{\measbody}{W}

\label{sec:sumsquare}
Our goal now is to prove \pref{thm:sos-general}. For a general seim-norm $N$ on $\R^n$, directly applying \pref{thm:l2-sparsification} with the parameters $p = 2$, $K = \sqrt{n}$, $S = 1$ leads to the unimpressive bound of $O(\e^{-2}n^2\log^5(n/\e))$. Towards improving this, we start by obtaining a (dimension-dependent) improvement on the shift lemma (\pref{lem:shift}).

\begin{lemma}\label{lem:gen-l2-shift}
   Suppose $\sudnorm$ is a norm on $\R^n$ and $p \in [1,2]$.
   Let $\mu$ denote the probability measure $\mu$ on $\R^n$ satisfying
   \[
      d\mu(x) \propto \exp(-\sudnorm(x)^p)\,dx\,.
   \]
   Then for any measureable ${\measbody} \subseteq \R^n$ and $z \in \R^n$, $\delta > 0$, we have
   \begin{equation}\label{eq:shift-measure2}
      \mu({\measbody}+z) \geq e^{-\delta n/p} e^{- \left(\frac{1+\delta}{\delta}\right)^{p-1} \sudnorm(z)^p} \mu({\measbody})\,.
   \end{equation}
\end{lemma}

\begin{proof}
   For any $\delta > 0$ and $x,z \in \R^n$, it holds that
   \[
      \sudnorm(x+z)^p \leq (\sudnorm(x)+\sudnorm(z))^p \leq (1+\delta)^{p-1} \sudnorm(x)^p + \left(\tfrac{1+\delta}{\delta}\right)^{p-1} \sudnorm(z)^p,
   \]
   therefore
   \begin{align*}
      \int_{\measbody} \exp\left(-\sudnorm(x+z)^p\right)\,dx
                 &\geq e^{-\left(\tfrac{1+\delta}{\delta}\right)^{p-1} \sudnorm(z)^p} \int_{\measbody} \exp\left(-(1+\delta)^{p-1} \sudnorm(x)^p\right)\,dx \\
                 &=\left(1+\delta\right)^{-(p-1)/p \cdot n} e^{-\left(\tfrac{1+\delta}{\delta}\right)^{p-1} \sudnorm(z)^p} \int_{\measbody} \exp(-\sudnorm(x)^p)\,dx \\
                 &\geq e^{-\delta n/p} e^{-\left(\tfrac{1+\delta}{\delta}\right)^{p-1} \sudnorm(z)^p}\int_{\measbody} \exp(-\sudnorm(x)^p)\,dx\,.
   \end{align*}
   To complete the proof, observe that
   \[
      \mu({\measbody} + z) = \frac{\int_{\measbody} \exp\left(-\sudnorm(x+z)^p\right)\,dx}{\int_{\measbody} \exp\left(-\sudnorm(x)^p\right)\,dx} \mu({\measbody})\,.\qedhere
   \]
\end{proof}

\begin{lemma}\label{lem:gen-l2-cover}
   Suppose $\sudnorm$ and $\hatsudnorm$ are semi-norms on $\R^n$ with $\ker(\sudnorm) \subseteq \ker(\hatsudnorm)$.
   Denote the probability measure $\mu$ on $\ker(\sudnorm)^{\perp}$ so that 
   \[
      d\mu(x) \propto \exp(-\sudnorm(x)^p)\,dx\,.
   \]
   Then for any $\e > 0$,
   \[
      \left(\log \cov(B_\sudnorm, \hatsudnorm, \e)\right)^{1/2} \lesssim \left(\frac{\lambda}{\e}\right)^{p/2} + \sqrt{\frac{\lambda}{\e}} n^{\frac12 - \frac{1}{2p}} \,,
   \]
   where
   \[
      \lambda = \int \hatsudnorm(x) \,d\mu(x)\,.
   \]
\end{lemma}

\begin{proof}
   By scaling $\hatsudnorm$, we may assume that $\e = 1$.
Suppose now that $x_1,\ldots,x_M \in B_{\sudnorm}$ and the balls $x_1+ B_{\hatsudnorm},\ldots,x_M+ B_{\hatsudnorm}$ are pairwise disjoint.
To establish an upper bound on $M$, let $\lambda > 0$ be a number
we will choose later and write
\begin{align*}
   1 \geq \mu\left(\bigcup_{j=1}^M \lambda (x_j + B_{\hatsudnorm})\right)  &= \sum_{j=1}^M \mu\left(\lambda x_j + \lambda B_{\hatsudnorm}\right) \\
                                                               &\stackrel{\mathclap{\eqref{eq:shift-measure2}}}{\geq}\ 
                                                               e^{-\delta n/p} \sum_{j=1}^M e^{- \lambda^p \left(\tfrac{1+\delta}{\delta}\right)^{p-1} \sudnorm(x_j)^p} \mu(\lambda B_{\hatsudnorm}) \\
                                                               &\geq M e^{-\delta n/p} e^{- \lambda^p \left(\tfrac{1+\delta}{\delta}\right)^{p-1}}\mu(\lambda B_{\hatsudnorm})\,,
\end{align*}
where the last inequality uses $x_1,\ldots,x_M \in B_{\sudnorm}$.
Now choose $\lambda \seteq 2 \int \hatsudnorm(x)\, d\mu(x)$
so that Markov's inequality gives
\[
   \mu(\lambda B_{\hatsudnorm}) = \mu\left(\{x : \hatsudnorm(x) \leq \lambda \} \right) \geq 1/2\,,
\]
yielding the upper bound
\[
   \log (M/2) \leq \frac{\delta n}{p} + \lambda^p \left(\tfrac{1+\delta}{\delta}\right)^{p-1}\,.
\]
Choosing $\delta \seteq \lambda/n^{1/p}$ and using $(1 + 1/\delta)^{p-1} \lesssim 1 + \delta^{-(p-1)}$ gives
\[
   \log (M/2) \lesssim \lambda n^{1-1/p} + \lambda^p\,.\qedhere
\]
\end{proof}

\subsubsection{Entropy estimate}

We will work again in the setting of \pref{def:entropy-objects}.

\begin{lemma}\label{lem:l2-chaining}
   Consider \pref{def:entropy-objects} and additionally
   \begin{align*}
      \kappa &\seteq \max_{x \in B_N} \cN^{\infty}(x)\,, \\
      \lambda &\seteq \psi_n \log(M) \mathsf{w}_{\infty}\,.
   \end{align*}
   Then it holds that
   \begin{equation}\label{eq:l2-chaining} 
      \gamma_2(B_{\bnorm}, d) \lesssim \left(\kappa^{\frac{p-1}{2}} \lambda^{\frac12} n^{\frac12 - \frac{1}{2p}} + n^{-1} \kappa^{\frac{p}{2}} + \lambda^{\frac{p}{2}} \log n\right) \sqrt{\Lambda_{B_{\bnorm}}}\,.
   \end{equation}
\end{lemma}

\begin{proof}
   Since both sides of \eqref{eq:l2-chaining} scale linearly in the values $\{\cN_j^p\}$, we may assume that
   \begin{equation}\label{eq:l2-scale}
      \Lambda_{B_{\bnorm}} = \max_{x \in B_{\bnorm}} \sum_{j=1}^M \cN_j(x)^p = 1\,,
    \end{equation}
    and then \pref{lem:inside0} gives
\begin{equation}\label{eq:l2-inside}
   d(x,y) \leq 2 \cN^{\infty}(x-y)^{p/2}\,,\qquad x,y \in B_{\bnorm}\,.
\end{equation}
This yields the comparisons
\begin{equation}\label{eq:l2-pre-dud2}
   \cov(B_{\bnorm}, d, r) \leq \cov(B_{\bnorm}, \cN^{\infty}, (r/2)^{2/p}) \leq \cov(B_{\cN^{\infty}}, \cN^{\infty}, \tfrac{(r/2)^{2/p}}{\kappa}) 
   \leq \left(\frac{2 \kappa}{(r/2)^{2/p}}\right)^n\,,
\end{equation}
where the second inequality uses the definition of $\kappa$, 
and the final inequality follows from \pref{lem:norm-cover}.

In particular, we have
\begin{equation}\label{eq:pre-dud3}
   \int_{0}^{\kappa^{p/2}/n^2} \sqrt{\log \cov(B_{\cN^{\infty}}, \cN^{\infty}, (r/2)^{p/2})}\,dr
   \lesssim \sqrt{n} \int_0^{\kappa^{p/2}/n^2} \sqrt{\log \frac{\kappa}{r^{2/p}}}\,dr
   \lesssim n^{-3/2} \kappa^{p/2} \log n\,.
\end{equation}
\pref{lem:gen-l2-cover} with $\sudnorm=\bnorm, \hatsudnorm=\cN^{\infty}$ asserts that
\begin{equation}
\label{eq:l2-cov}
(\log \cov(B_{\bnorm}, \cN^{\infty}, (r/2)^{2/p})^{1/2} 
\lesssim \left(\frac{\lambda_0}{(r/2)^{2/p}}\right)^{p/2} + \sqrt{\frac{\lambda_0}{(r/2)^{2/p}}} n^{\frac12 - \frac{1}{2p}}
   \lesssim \frac{\lambda_0^{p/2}}{r} + \frac{\lambda_0^{1/2}}{r^{1/p}} n^{\frac12-\frac{1}{2p}}\,,
\end{equation}
where $\lambda_0 \seteq \int \cN^{\infty}(x) \,d\mu(x)$.

Dudley's entropy bound \eqref{eq:dudley-v2} in conjunction with \eqref{eq:l2-pre-dud2} gives
\[
   \gamma_2(B_n,d) \lesssim \int_0^{\infty} \sqrt{\log \cov(B_{\bnorm},d,r)}\,dr
   \lesssim \int_0^{(2 \kappa)^{p/2}} \sqrt{\log \cov(B_{\bnorm}, \cN^{\infty}, (r/2)^{2/p})}\,dr\,,
\]
where we have used the fact that $\log \cov(B_{\bnorm},\cN^{\infty},r)=0$ for $r \geq \kappa$.
Now using \eqref{eq:pre-dud3} and \eqref{eq:l2-cov} yields
\begin{align*}
   \gamma_2(B_{\bnorm},d) &\lesssim \frac{\kappa^{p/2}}{n} + \lambda_0^{p/2} \int_{\kappa^{p/2}/n^2}^{(2\kappa)^{p/2}} \frac{1}{r}\,dr
   + \lambda_0^{1/2} n^{\frac12-\frac{1}{2p}} \int_{\kappa^{p/2}/n^2}^{(2\kappa)^{p/2}} \frac{1}{r^{1/p}}\,dr
   \\
                   &\lesssim \frac{\kappa^{p/2}}{n} + \lambda_0^{p/2} \log n + \lambda_0^{1/2} n^{\frac12-\frac{1}{2p}} \kappa^{\frac{p-1}{2}}\,.
\end{align*}
To conclude, recall that \pref{lem:tailbound} gives the estimate
\[
   \lambda_0 \lesssim \psi_n \log(M) \mathsf{w}_{\infty} = \lambda\,.\qedhere
\]
\end{proof}

\subsubsection{Sparsification}

We obtain our result for sparsifying sums of $p$th powers of arbitrary norms. 

\begin{theorem}\label{thm:l2ofnorms}
Suppose $1 \leq p \leq 2$ and
${N}_1,\ldots,{N}_m$ are semi-norms on $\R^n$. Denote 
\[
   N(x) \seteq ({N}_1(x)^p+\cdots+{N}_m(x)^p)^{1/p}\,.
\]
Then there is a weight vector $w \in \R_+^m$ with
\[
   |\supp(w)| \lesssim \frac{n^{2-1/p} \psi_n \log(n/\e)}{\e^2} + \frac{n (\psi_n \log (n/\e))^p (\log n)^2}{\e^2}\,,
\]
and such that
   \[
      \left|N(x)^p-\sum_{j=1}^m w_j N_j(x)^p\right| \leq \e N(x)^p\,,\qquad \forall x\in \R^n.
   \]
\end{theorem}

\begin{proof}[Proof of \pref{thm:l2ofnorms}]
   Let $U : \R^n \to \R^n$ be the transformation guaranteed by \pref{lem:block-lewis} (with $p=q$).
   In particular, there are weights $\alpha_1,\ldots,\alpha_m \geq 0$
   such that $\alpha_1^p+\cdots+\alpha_m^p \leq n$, and
\begin{equation}\label{eq:Njalpha}
   N_j(x)^p \leq \alpha_j^p N(x)^p\,,\quad \forall x\in \R^n\,.
\end{equation}

Let $\mu$ denote the measure with density $d\mu(x) \propto e^{-N(x)^p} dx$, and define for $i \in [m]$,
\begin{align*}
   \tau_i &\seteq \int {N}_i(x)^p\,d\mu(x) \\
   \rho_i &\seteq \frac{\tau_i+\alpha_i^p}{\sum_{i=1}^m (\tau_i + \alpha_i^p)}\,.
\end{align*}
By \pref{lem:int-Nz}, we have

\begin{equation}
   \label{eq:l2-normalized}
   \tau_1 + \cdots + \tau_m = \int N(x)^p\,d\mu(x) = \frac{n}{p}\,.
\end{equation}

We will apply \pref{lem:l2-chaining} with $\bnorm=N$.
Given $M \geq 1$ and $\nu \in [m]^M$, define
\begin{align*}
   \cN_j(x) &\seteq \frac{N_{\nu_j}(x)}{(M \rho_{\nu_j})^{1/p}}\,,\quad j=1,\ldots,M\,, \\
   \f_i(x) &\seteq N_i(x)^p\,,\qquad i=1,\ldots,m\,,
\end{align*}
so that $d_{\rho,\nu}(x,y)=d(x,y)$, where $d$ is the distance from \eqref{def:entropy-objects}.
Observe that 
\[
   \frac{N_i(x)^p}{\rho_i} \leq \sum_{i=1}^m (\tau_i + \alpha_i^p) \frac{N_i(x)^p}{\alpha_i^p} \stackrel{\eqref{eq:Njalpha}}{\leq} 3n N(x)^p\,, \quad \forall x \in \R^n\,, i=1,\ldots,m\,,
\]
and moreover
\[
   \int \cN_j(x)^p\,d\mu(x) = \frac{1}{M \rho_{\nu_j}} \int N_{\nu_j}(x)^p\,d\mu(x) = \frac{\tau_{\nu_j}}{M \rho_{\nu_j}} \leq \frac{3n}{M}\,, \quad j=1,\ldots,M\,.
\]
Therefore $\mathsf{w}_{\infty} \leq (3n/M)^{1/p}$, and
$\kappa\lesssim \left(\frac{n}{M}\right)^{1/p}$ and $\lambda \lesssim \left(\frac{n}{M}\right)^{1/p} \psi_n \log M$ in \pref{lem:l2-chaining}.

It follows that
\begin{align*}
   \gamma_2(B_N, d) &\lesssim \left(\frac{n}{M}\right)^{1/2} \left(n^{\frac12 - \frac{1}{2p}} (\psi_n \log M)^{1/2} + (\psi_n \log M)^{p/2} \log n \right) \left(\max_{N(x) \leq 1} \sum_{j=1}^M \cN_j(x)^p\right)^{1/2} \\
   &\lesssim \e \left(\max_{N(x) \leq 1} \sum_{j=1}^M \cN_j(x)^p\right)^{1/2},
\end{align*}
for some choice of $M$ satisfying
\[
   M \lesssim \frac{n^{2-1/p} \psi_n \log(n/\e)}{\e^2} + \frac{n (\psi_n \log (n/\e))^p (\log n)^2}{\e^2}\,.
\]
With this, \pref{lem:sparsification-meta} yields
\[
   \E_{\bm{\nu}} \max_{x \in B_N} \left|N(x)^p - \frac{1}{M} \sum_{j \in [M]} \frac{N_{\bm{\nu}_j}(x)^p}{\rho_{\bm{\nu}_j}} \right| \lesssim \e.
\]
Thus there exists a choice of $\nu \in [m]^M$ satisfying the bound, and the claim follows.
\end{proof}

\subsection{Sums of squares of $\ell_p$ norms}

Our goal now is to prove \pref{thm:sos-lp}.
We will require the following chaining estimate.

\begin{theorem}[{\cite[Lemma 2.12]{Lee23}}]
	\label{thm:lee22}
	Suppose $\cN_1,\ldots,\cN_M$ are semi-norms on $\R^n$, and
	define
	\begin{equation}\label{eq:var-distance}
		d(x,y) \seteq \left(\sum_{j=1}^M \left(\cN_j(x)^2 - \cN_j(y)^2\right)^2\right)^{1/2}
	\end{equation}
	Then for any set $T \subseteq B_2^n$, it holds that
	\[
	\gamma_2(T,d) \lesssim \sup_{x \in T} \left(\sum_{j=1}^M \cN_j(x)^4\right)^{1/2} + 
	\left(\kappa + \lambda \sqrt{\log n}\right) \sup_{x \in T} \left(\sum_{j=1}^M \cN_j(x)^2\right)^{1/2}\,,
	\]
	where
	\begin{align*}
		\kappa &\seteq \E \max_{j \in [M]} \cN_{j}(\bm{g}) \\
		\lambda &\seteq \max_{j \in [M]} \E \cN_{j}(\bm{g})\,,
	\end{align*}
	and $\bm{g}$ is a standard $n$-dimensional Gaussian.
\end{theorem}

\begin{proof}[Proof of \pref{thm:sos-lp}]
   By scaling the matrices $\{A_j\}$ suitably, we may assume that
   \begin{equation}\label{eq:Aj-equiv}
   \|A_j x\|_{p_j} \leq N_j(x) \leq K \|A_j x\|_{p_j}\,,\qquad  \forall  x \in \R^n, j=1,\ldots,m\,.
\end{equation}
	Define $A : \R^n \to \R^k$ for $k \seteq n_1 + \dots + n_m$ by $A(x) = A_1(x) \oplus \cdots \oplus A_m(x)$ and let $S_1 \cup \cdots \cup S_m = [k]$
	be the natural partition.
   Let $U \in \cL(\R^n)$ be the transformation established in \pref{lem:lp-lewis},
   and denote $\tau_j \seteq \alpha_j(U)^2$ for $j=1,\ldots,m$.
   Note that \pref{lem:lp-lewis}\pref{enumlew:a} ensures $\|\tau\|_1 \leq n$.

   \smallskip

	For $j \in [m]$, define
	\begin{align*}
		\rho_j &\seteq \tau_j/\|\tau\|_1\,, \\
		\varphi_j(x) &\seteq N_j(x)^2\,.
	\end{align*}
	Then using \pref{lem:sparsification-meta}, our goal is to bound $\gamma_2(B_N, d_{\rho,\nu})$ for any $\nu \in [m]^M$.
	
	To this end, define
	\begin{align*}
		\hat{N}_j(x) &\seteq \|A_j U x\|_{p_j} \,,\qquad j=1,\ldots,m\,, \\
		\hat{N}(x) &\seteq \left(\hat{N}_1(x)^2 + \cdots + \hat{N}_m(x)^2\right)^{1/2},
	\end{align*}

   and then the second inequality in \pref{lem:lp-lewis}\pref{enumlew:b} implies
	\[
	\|x\|_2 \leq \hat{N}(x)\,,
	\]
	hence $B_{\hat{N}} \subseteq B_2^n$, and $\hat{N}_j(x) \leq \tau_j \|x\|_2^2$ for all $x \in \R^n, j \in [m]$.
	
   \smallskip

	Fix $\nu \in [m]^M$ and 
	let us finally define, for $i \in [M]$,
	\begin{align*}
		\cN_i(x) &\seteq \frac{\hat{N}_{\nu_i}(x)}{\sqrt{\rho_{\nu_i}}}\,.
	\end{align*}
	Let $d$ be the corresponding distance given by \eqref{eq:var-distance}.
	Then by construction, we have
   \begin{equation}\label{eq:d-compare}
	d_{\rho,\nu}(x,y) \leq K^2 \frac{d(U^{-1} x, U^{-1} y)}{M}\,,
\end{equation}
   where the inequality uses \eqref{eq:Aj-equiv}.

	Since $B_{\hat{N}} \subseteq B_2^n$, we can apply \pref{thm:lee22} with $T = B_{\hat{N}}$.
   Let us note first that, by \pref{lem:lp-lewis}\pref{enumlew:b} we have $\hat{N}_j(x)^2 \leq \tau_j \|x\|_2^2 \leq \tau_j \hat{N}(x)^2$.
   Therefore
	$\cN_j(x)^2 \leq \|\tau\|_1 \hat{N}(x)^2$, and
	\begin{equation}\label{eq:l4-bnd}
		\max_{\hat{N}(x) \leq 1} \sum_{j=1}^M \cN_j(x)^4 \leq \|\tau\|_1 \max_{\hat{N}(x) \leq 1} \sum_{j=1}^M \cN_j(x)^2\,.
	\end{equation}
	
	Moreover, it holds that
   \begin{align*}
      \E \hat{N}_j(\bm{g}) \leq \left(\E \hat{N}_j(\bm{g})^{p_j} \right)^{1/p_j} &= \left(\sum_{i \in S_j} \E |\langle a_i, U \bm{g}\rangle|^{p_j}\right)^{1/p_j}  \\
                                                                       &=  \left(\sum_{i \in S_j} \|U a_i\|_2^{p_j} \E |\bm{g}_1|^{p_j}\right)^{1/p_j} \\
                                                                       &\lesssim \sqrt{p_j} \left(\sum_{i \in S_j} \|U a_i\|_2^{p_j} \right)^{1/p_j}
                                                                       = \sqrt{p_j}\ \alpha_j(U) 
                                                                       \leq \sqrt{p \tau_j}\,,
   \end{align*}
   where the first inequality uses the fact that $(\E[|\bm{g}_1|^{p}])^{1/p} \lesssim \sqrt{p}$ for any $p \geq 1$, and the second uses $p_j \le p$.

   Recall that $\hat{N}_j(x)^2 \leq \tau_j \|x\|_2^2$, and therefore the Gaussian concentration inequality (\pref{thm:gaussian-concentration})
   implies that
	\[
      \Pr\left(\hat{N}_{j}(\bm{g}) > \E \hat{N}_{j}(\bm{g}) + t \sqrt{\tau_j}\right) \leq e^{-t^2/2}\,.
	\]
	In particular, this gives
	\[
      \E \max_{i \in [M]} \frac{\hat{N}_{\nu_i}(\bm{g})}{\sqrt{\rho_{\nu_i}}} \lesssim \sqrt{\|\tau\|_1 \log M}\,.
	\]
	We conclude that
	\begin{align*}
      \lambda &= \max_{j \in [M]} \E \cN_j(\bm{g}) \leq \sqrt{p \|\tau\|_1}\,, \\
      \kappa &= \E \max_{j \in [M]} \cN_j(\bm{g}) \lesssim \sqrt{\|\tau\|_1 (p + \log M)}\,.
	\end{align*}
	Combining these with \eqref{eq:l4-bnd} and $\|\tau\|_1 \leq n$, \pref{thm:lee22} yields
	\begin{equation}\label{eq:gamma2-hat}
      \gamma_2(B_{\hat{N}}, d) \lesssim \sqrt{n} \left(1 + \sqrt{p + \log M} + \sqrt{p \log n}\right) \left(\max_{\hat{N}(x) \leq 1} \sum_{j=1}^M \cN_j(x)^2\right)^{1/2}\,.
	\end{equation}

   Note that
   \[
      \sum_{j=1}^M \cN_j(x)^2 = \sum_{j=1}^M \frac{\hat{N}_{\nu_i}(x)^2}{\rho_{\nu_i}}
      \stackrel{\eqref{eq:Aj-equiv}}{\leq} \sum_{j=1}^M \frac{N_{\nu_i}(U^{-1} x)^2}{\rho_{\nu_i}}
      = M \tilde{F}_{\rho,\nu}(U^{-1} x)\,.
   \]
   And since \eqref{eq:Aj-equiv} implies $N(U^{-1} x) \leq K \hat{N}(x)$, 
   we have
	\[
	\max_{\hat{N}(x) \leq 1} \sum_{j=1}^M \cN_j(x)^2 \leq K^2 M \max_{N(U^{-1} x) \leq 1} \tilde{F}_{\rho,\nu}(U^{-1} x) \leq K^2 M
	\max_{N(x) \leq 1} \tilde{F}_{\rho,\nu}(x)\,.
	\]
	Using this together with \eqref{eq:gamma2-hat} gives
	\[
      \gamma_2(B_{\hat{N}}, d) \lesssim K M^{1/2} \sqrt{n (p \log n + \log M)} \left(\max_{N(x) \leq 1} \tilde{F}_{\rho,\nu}(x)\right)^{1/2}.
	\]
   Combining this with \eqref{eq:d-compare}, we conclude that
	\[
      \gamma_2(B_{N}, d_{\rho,\nu}) \lesssim K^3 M^{-1/2} \sqrt{n (p \log n + \log M)} \left(\max_{N(x) \leq 1} \tilde{F}_{\rho,\nu}(x)\right)^{1/2}.
	\]

	Now choose $M \asymp \frac{K^6 p n \log (n/\e)}{\e^2}$ and apply \pref{lem:sparsification-meta} to obtain
	\[
	\E_{\bm{\nu}} \max_{N(x) \leq 1} \left|N(x)^2 - \frac{1}{M} \sum_{j=1}^M \frac{N_{\bm{\nu}_j}(x)^2}{\rho_{\bm{\nu}_j}}\right|\leq \e\,,
	\]
	completing the proof.
\end{proof}

\section*{Acknowledgments}

We thank anonymous reviewers for several helpful comments.
James R. Lee is supported in part by NSF CCF-2007079 and a Simons Investigator Award. 
Yang P. Liu is partially supported by the Google Research Fellowship and NSF DMS-1926686.
Aaron Sidford is supported in part by a Microsoft Research Faculty Fellowship, NSF CCF-1844855, NSF CCF-1955039, a PayPal research award, and a Sloan Research Fellowship.

\bibliographystyle{alpha}
\bibliography{norm-sparsify}

\end{document}